\documentclass[a4paper,11pt]{article}
\pdfoutput=1
\usepackage{jheppub2}
\usepackage{graphicx}
\usepackage{amsmath}
\usepackage[T1]{fontenc} % if needed

\newcommand{\tr}{\mathop{\rm tr}}
\newcommand{\plb}[3]{{\it Phys.\ Lett.\ }{\bf B #1} (#2) #3}
\newcommand{\mpla}[3]{{\it Mod.\ Phys.\ Lett.\ }{\bf A #1} (#2) #3}
\newcommand{\npb}[3]{{\it Nucl.\ Phys.\ }{\bf B #1} (#2) #3}
\newcommand{\jhep}[3]{{\it J. High Energy Phys.\ }{\bf #1} (#2) #3}
\newcommand{\prl}[3]{{\it Phys.\ Rev.\ Lett.\ }{\bf #1} (#2) #3}
\newcommand{\ijmpa}[3]{{\it Int.\ J.\ Mod.\ Phys.\ }{\bf A #1} (#2) #3}
\newcommand{\PR}[3]{{\it Phys.\ Rept.\ }{ #1} (#2) #3}
\newcommand{\hepth}[1]{\href{http://xxx.lanl.gov/abs/hep-th/#1}{\tt hep-th/#1}}
\newcommand{\hepph}[1]{\href{http://xxx.lanl.gov/abs/hep-ph/#1}{\tt hep-ph/#1}}

\title{Revisiting the S-matrix approach to the open superstring low energy effective lagrangian}

\author[a]{Luiz Antonio Barreiro,}
\author[b,1]{and Ricardo Medina\note{Corresponding author.}}

\affiliation[a]{Departamento de F{\'i}sica, UNESP,\\Rio Claro, S{\~a}o Paulo, Brazil}
\affiliation[b]{Instituto de Ci{\^e}ncias Exatas, Universidade Federal de Itajub\'{a},\\Itajub\'a, Minas Gerais, Brazil}

\emailAdd{luiz.a.barreiro@gmail.com}
\emailAdd{rmedina50@gmail.com}

\abstract{The conventional S-matrix approach to the (tree level) open string low energy effective lagrangian assumes that, in order to obtain all its bosonic ${\alpha'}^N$ order terms, it is necessary to know the open string (tree level) $(N+2)$-point amplitude of massless bosons, at least expanded at that order in $\alpha'$. In this work we clarify that the previous claim is indeed valid for the bosonic open string, but for the supersymmetric one the situation is much more better than that: there are constraints in the kinematical bosonic terms of the amplitude (probably due to Spacetime Supersymmetry) such that a much lower open superstring $n$-point amplitude is needed to find all the ${\alpha'}^N$ order terms. In this `revisited' S-matrix approach we have checked that, at least up to ${\alpha'}^4$ order, using these kinematical constraints and only the known open superstring $4$-point amplitude, it is possible to determine all the bosonic terms  of the low energy effective lagrangian. The sort of results that we obtain seem to agree completely with the ones achieved by the method of BPS configurations, proposed about ten years ago. By means of the KLT relations, our results can be mapped to the NS-NS sector of the low energy effective lagrangian of the type II string theories implying that there one can also find kinematical constraints in the $N$-point amplitudes and that important informations can be inferred, at least up to ${\alpha'}^4$ order, by only using the (tree level) 4-point amplitude.}

\begin{document}
\maketitle
\flushbottom

\section{Introduction}

\label{Introduction}

The low energy effective lagrangian in String Theory is a very old subject, dating back to the beginnings of the theory, when it became clear that, in the limit of very low energies, it reproduces the (tree level) scattering amplitudes of General Relativity and Yang-Mills theories. More precisely, it was known that it adds $\alpha'$ correction terms to the lagrangians of these theories \cite{Scherk1}, where $\alpha'$ is the string fundamental constant.\\
In the mid-eighties , after the understanding that  consistency conditions of String Theory (quantum conformal invariance) demand that the extrema of the low energy effective action be related to the zeroes of the Sigma model beta function \cite{Callan1, Tseytlin2}, interest arised in calculating exactly the first $\alpha'$ correction terms in the low energy effective actions of superstrings \cite{Gross1, Tseytlin2}. One of the amazing results that was found by that time is that the Born-Infeld lagrangian (and its corresponding supersymmetric version, in the case of superstrings) is the low energy effective lagrangian for abelian open strings, as long as $F_{\mu \nu}$ is kept constant \cite{Fradkin1, Tseytlin2}. This result was reproduced afterwards as well, independently, in  \cite{Abouelsaood1, Bergshoeff3, Leigh1}\footnote{Recently, using a pure spinor framework, the abelian supersymmetric Born-Infeld theory has been formulated by means of a polynomial lagrangian which, besides the free term, contains only 4-point interaction terms\cite{Cederwall3}.}. \\
Unfortunately, in the nonabelian case there is not such a nice result. There does exist a `Symmetrized Trace' proposal for the nonabelian Born-Infeld lagrangian \cite{Tseytlin1}, and this proposal indeed works at ${\alpha'}^2$ order, but is clear from the $[ D , D ] \cdot = [F, \cdot ]$ identity (see eq. (\ref{group4})), that  covariant derivative terms are as important as the ones without derivatives and, therefore, they cannot be ignored as it happens in the abelian theory. Also, it is clear that the Symmetrized Trace prescription does not work already at ${\alpha'}^3$ order since the usual Born-Infeld lagrangian only contains even powers of $\alpha'$ and it is known that the ${\alpha'}^3$ terms are not zero in the nonabelian low energy effective action \cite{Kitazawa1, Bilal1, Refolli1, Koerber1, DeRoo2, Brandt1}.\\
Another approach that could be used to obtain the low energy effective action is $\kappa$-symmetry. Although this approach works in the abelian case, it was seen in \cite{Bergshoeff0} that in the nonabelian case it fails at ${\alpha'}^3$ order. So, for the nonabelian case, besides the results obtained in \cite{Chandia1} and \cite{Barreiro1}, there does not exist an all $\alpha'$ order result (like the Born-Infeld one, in the abelian case)\footnote{The results in \cite{Chandia1} and \cite{Barreiro1} have been obtained using $4$ and $5$-point amplitudes and, therefore, only determine terms of the nonabelian lagrangian which are sensible to those $n$-point amplitudes. For example, the ${\alpha'}^4 F^6$ terms are not present in these references.}'\footnote{A later and independent calculation of the $D^{2n}F^4$ terms, at every $\alpha'$ order, was done in \cite{Ehsan1}, also considering the open superstring  4-point amplitude.}.\\
In the mid-nineties, after the discovery that $D$-branes participate in String Theory as non-perturbative objects \cite{Polchinski0} and that $D$-brane effective actions can be obtained equivalently by means of low energy limit of open superstring interactions \cite{Witten1}, new interest arised in the calculation of terms of the low energy effective lagrangian (see \cite{Wyllard2, Andreev2, Koerber1, Koerber2, DeRoo1}, for example). Before the work in \cite{Brandt1}, the S-matrix approach for the $\alpha'$ correction terms in the low energy effective action had been only used with knowledge of $4$-point amplitudes \cite{Bilal1}, which was already insufficient to determine all the ${\alpha'}^3$ terms. Alternative methods, mainly based on Supersymmetry, arised to obtain the complete list of nonabelian terms at each order in $\alpha'$, whenever possible \cite{Refolli1, Fosse1, DeRoo2, Grasso1, Cederwall1, Cederwall2, Drummond1, Grasso2, Movshev1, Howe1}. In some cases these methods obtained the complete superinvariant at a given order in $\alpha'$, either in component fields \cite{Cederwall1, Cederwall2, Movshev1, Howe1} or in terms of superfields \cite{Grasso1, Drummond1} (in the case of D=4). In some other cases these methods could only obtain some of the bosonic and fermionic terms, like in \cite{DeRoo2}, but they were capable of proving that there is a unique superinvariant at that $\alpha'$ order. The thing is that, in order to go to higher orders in $\alpha'$, these Supersymmetry based methods usually require to know, at some moment, all the bosonic and fermionic terms of the lagrangian at a lower order in $\alpha'$, and this demands an enormous effort. In order to get contact with the open superstring low energy effective lagrangian (OSLEEL) all these methods required at some moment information from the (tree level) open superstring $4$-point amplitude.\\
There is an additional alternative method that was not mentioned in the previous paragraph because it is not directly related to Supersymmetry, but it has the virtue of obtaining the bosonic terms of the OSLEEL correctly \cite{Koerber1} without having to deal with the fermionic ones. Besides the ${\alpha'}^2$ terms, the results obtained by this method have been checked at ${\alpha'}^3$ order \cite{Koerber3, Brandt1} and there is evidence, in the self consistency of its calculations and also when considering its abelian limit, that its bosonic ${\alpha'}^4$ terms are also correct \cite{Koerber2}. This method consists in working with BPS configurations in the deformed Yang-Mills lagrangian \cite{Koerber4} and it is due to Koerber and Sevrin. \\
\noindent Actually, a complete low energy description of $D$-branes not only depends on the Dirac-Born-Infeld lagrangian (which already describes interactions between open and closed strings), but also on the Wess-Zumino lagrangian (which considers the interactions between open strings and the Ramnod-Ramond sector of closed strings; see, for example, the introduction of \cite{Ehsan0} for a recent review on the subject and references \cite{Ehsan01, Ehsan02} for several attempts with non-BPS branes.). A revival on S-matrix calculations to determine the $D$-brane low energy effective lagrangian terms (besides the pure gauge field ones) and also to determine the Wess-Zumino lagrangian terms is been going on this year \cite{Ehsan2, Ehsan3}.\\
\noindent In this work we will focus only in gauge boson string interactions. We propose a `revisited' S-matrix approach to obtain the bosonic terms of the (tree level) OSLEEL. We have called it `revisited' in order to distinguish it from the `conventional' one because it consists in a convenient way of dealing with the S-matrix calculations. The procedure takes into account an important kinematical constraint that arises in the calculations of the scattering amplitudes of open superstrings, namely, the absence of $(\zeta \cdot k)^N$ terms in the $N$-point scattering amplitude of gauge bosons. This constraint is not expected to be present in the corresponding calculations with open bosonic strings\footnote{As will be seen on this work, it is a confirmed fact that the constraint does not appear in $3$ and $4$-point amplitudes and, based on a general argument, it is not expected to appear in higher $N$-point amplitudes, as well.}. It leads to a set of conditions that the coefficients of the OSLEEL should obey, reducing enormously the number of unknown ones present in that lagrangian at a given order in $\alpha'$. These conditions are such that, at least up to ${\alpha'}^4$ order, all bosonic terms of the OSLEEL can be determined by purely using the known open superstring $4$-point amplitude. No higher $N$-point amplitude from Open Superstring Theory is required\footnote{Notice that, since the bosonic terms at ${\alpha'}^4$ order have the general form $F^6 + D^2 F^5 + D^4 F^4$, in order to find explicitly those terms the `conventional' S-matrix method would require an open superstring $6$-point amplitude, expanded at that $\alpha'$ order.} and this is the main remark of our revisited S-matrix method. This is the same type of result obtained by the method of BPS configurations \cite{Koerber2}. \\
The structure of this work is as follows. In section \ref{Review} we do a brief review of the conventional S-matrix approach to the low energy effective lagrangian in Open String Theory. We present enough material to claim that the way this method works, when considered at the level of the bosonic terms of the low energy effective lagrangian, it apparently makes no difference between the bosonic and the supersymmetric open string calculations. In section \ref{A new} we present the `revisited' S-matrix method approach to the OSLEEL, explaining its main difference with the conventional one. As concrete examples of the `revisited' S-matrix method, in sections \ref{Applying} and \ref{Low} we apply it to the determination of the OSLEEL up to ${\alpha'}^4$ order\footnote{We remind the reader that ${\alpha'}^4$ is the highest order for which the nonabelian bosonic terms of the OSLEEL have been completely obtained explicitly up to this moment \cite{Koerber2}. In ref. \cite{Howe1}, using a spinorial cohomology approach (which is related to the pure spinor formalism in D=10), a proof has been given for the existence of ${\alpha'}^4$ superinvariant as a deformation of the ${\alpha'}^2$ one and an algorithm is provided to find all its terms, but there it is not given an explicit expression of the bosonic ${\alpha'}^4$ terms.\\
In the case of the abelian theory at ${\alpha'}^4$ order, a supersymmetric version of it has been obtained in \cite{DeRoo3}.}. In section \ref{Summary} we end this work by giving a brief summary of our work and also mentioning future contributions that will come out as results that arose from our present investigation.  We also comment there on the implications that our results have for the NS-NS sector of the type II String Theories once one considers the KLT relations \cite{Kawai1}.\\
The main body of this work is complemented with a series of appendices which are important to support claims and intermediate calculations which were omitted on it.\\
Through out this work all our scattering amplitude calculations are tree level ones and, therefore, the terms of the low energy effective lagrangian that we deal with are only `single' trace ones. Every time that we refer to a scattering amplitude it is understood that it is a tree level one (unless explicitly specified something different).

\section{Brief review of gauge boson scattering amplitudes and low energy effective theory in Open String Theory}

\label{Review}

In this section we briefly review the conventional S-matrix approach to the low energy effective lagrangian in Open Superstring Theory.

\subsection{Tree level gauge boson interactions in Open String Theory}

\label{Tree level}

Tree level scattering amplitudes of bosonic states in (non-abelian) Open Superstring Theory are given by a sum of (color ordered) subamplitudes \cite{Schwarz1}:
\begin{eqnarray}
{\cal A}_N & = & i (2 \pi)^{10} \delta^{10} (k_1 + \ldots +k_N ) \ \biggl[ \ \mbox{tr}(\lambda^{a_1} \lambda^{a_2} \ldots \lambda^{a_N}) \ A(1, 2, \ldots, N) +
\left( \begin{array}{c}
          \mbox{non-cyclic} \\
          \mbox{permutations}
          \end{array}
\right)  \ \biggr]   ,\nonumber \\ \ \
\label{N-point}
\end{eqnarray}
where the subamplitude $A(1, 2, \ldots, N)$ is given by an integrated vacuum average of vertex operators inserted on the boundary of a disk in the ordering $(1, 2, \ldots, N)$ \cite{Green1, Polchinski1}:
\begin{eqnarray}
 \hspace{-20pt} A(1, 2, \ldots, N) & = & g^{N-2} \ \int d \mu(z) \ <\hat{V}_1(z_1, k_1) \hat{V}_2(z_2, k_2) \ \ldots \hat{V}_N(z_N, k_N) > \ , \ \ N \ge 3 \ .
\label{A12N}
\end{eqnarray}
In (\ref{A12N}) $g$ is the open string coupling constant and $d \mu(z)$ is the $SL(2,R)$ invariant measure associated to the coordinates $z_1$, $z_2$, $\ldots$, $z_N$.\\
In formula (\ref{N-point}) ${\cal A}_N$ has been written as a sum of $(N-1)!$ contributions, each of them containing a trace color factor. In the case of Bosonic Open String Theory these factors terms are generally independent, but in the case of Supersymmetric Open String Theory the gauge group matrices $\lambda^{a}$'s are in the adjoint representation (see eq.(\ref{adjointm})) and this implies that in those $(N-1)!$ trace factors only half of them are independent, because in that representation it is valid that
\begin{eqnarray}
\mbox{tr}(\lambda^{a_1} \lambda^{a_2} \ldots \lambda^{a_N})= (-1)^N \mbox{tr}(\lambda^{a_N} \lambda^{a_{N-1}} \ldots \lambda^{a_1}) \ .
\label{lambdas}
\end{eqnarray}
This fact will be of importance in the results that we will present in section \ref{Low} (based on the calculations the we explain in Appendix \ref{Some details}). \\
In the case of the open superstring, described in the $F_1$ picture (old covariant approach) \cite{Green1}\footnote{The factors $\lambda$, which appears in (\ref{fermionic}), is a constant which depends on $\alpha'$ and may be explicitly determined by demanding unitarity relation for the $N$-point amplitude to be satisfied \cite{Polchinski1}.}
,
\begin{eqnarray}
\hat{V}(\zeta, k, z) = \lambda \ : ( \zeta \cdot \partial \hat{X}(z)  - i (2 \alpha') (\zeta \cdot \hat{\psi}) (k \cdot \hat{\psi}) ) \ e^{i k \cdot \hat{X}(z)}  : \ ,
\label{fermionic}
\end{eqnarray}
and after introducing convenient Grassmann variables $\theta_i$ and $\phi_j$, (\ref{A12N}) may be proved to become\footnote{In (\ref{ANfermionic}) it is only understood that  $\theta_{N-1}=\theta_{N}=0$. We will afterwards set $x_1=0$, $x_{N-1}=1$, $x_N= + \infty$. Formula (\ref{ANfermionic}) has been  taken from eq. (2.2) of \cite{Brandt1} and then a rescaling in the $\phi_i$ variables, by a factor of $(2 \alpha')^ {7/4}$, has been done.}
\begin{eqnarray}
A(1, 2, \ldots,N) & = & 2 \frac{g^{N-2}}{(2 \alpha')^2}
 (x_{N-1}-x_1)(x_{N}-x_1) \
 \int_0^{x_{N-1}} dx_{N-2} \int_0^{x_{N-2}} dx_{N-3} \ \dots  \int_0^{x_3} dx_2 \ \times \nonumber \\
&& {} \times \int d \theta_1 \ldots d \theta_{N-2} \prod_{p<q}^N (
 x_q - x_p - \theta_q \theta_p )^{2 \alpha' k_p \cdot k_q}\times  \int d \phi_1 \ldots d \phi_N
\nonumber \\
& &
\times \mbox{exp} \left(   \sum_{i \neq j}^N
\frac{  (2 \alpha')^1  (\theta_j-\theta_i) \phi_j (\zeta_j \cdot k_i) -1/2 \ (2 \alpha')^1  \phi_j \phi_i (\zeta_j \cdot \zeta_i) }{x_j-x_i-\theta_j \theta_i}    \right) \ .
\label{ANfermionic}
\end{eqnarray}
The subamplitude $A(1, \ldots,N)$, given in (\ref{ANfermionic}), satisfies important symmetries, such as (on-shell) gauge invariance,
cyclic symmetry and twisting (world-sheet parity) symmetry \cite{Schwarz1}. These symmetries provide a non trivial test that the explicit kinematical expression of $A(1, \ldots,N)$ should obey.\\
Recently, it has been proved that the set of all $N$-point color ordered subamplitudes, in bosonic as well as supersymmetric Open String Theory, can be expanded in a minimal basis of $(N-3)!$ subamplitudes \cite{Bjerrum1, Stieberger1}.\\
In the next subsection we will explicitly see that the open string coupling constant $g$ may be identified with the Yang-Mills coupling constant, $g_{YM}$.

\subsubsection{$3$ and $4$-point amplitudes}

\label{3 and 4}

On kinematical basis it may be argued that the 3-point amplitude, ${\cal A}_3$, for massless external states is zero, unless all three momenta are collinear. Along the formulas we presented in the previous subsection this can be seen in equation (\ref{N-point}), since for $N=3$ (and $k_1^2=k_2^2=k_3^2=0$) we have that $\delta^{10} (k_1 + k_2 +k_3)=0$ \footnote{Unless $k_i^{\mu}=0$ for one of the three states, which really  means that there are only two physical states at all.}.  \\
But the 3-point subamplitude, $A(1, 2, 3)$, is not zero. Its expression is important, for instance, in the determination of the low energy effective lagrangian.\\
Using formula (\ref{ANfermionic}) in the case of $N=3$ leads to \cite{Schwarz1}
\begin{eqnarray}
A(1,2,3) & = & 2 g \ \big[   \ (\zeta_1 \cdot k_2)(\zeta_2
\cdot \zeta_3) + (\zeta_2 \cdot k_3)(\zeta_3 \cdot \zeta_1) + (\zeta_3
\cdot k_1)(\zeta_1 \cdot \zeta_2) \   \big] \ ,
\label{A3fermionic}
\end{eqnarray}
that is, it agrees exactly with the corresponding Yang-Mills 3-point subamplitude, after identifying $g$ with $g_{YM}$.\\
Now, in the case of $N=4$, the expressions for the gauge boson subamplitude is given by \cite{Schwarz1}
\begin{eqnarray}
A(1,2,3,4) & = & 8 \ g^2 \ {\alpha'}^2 \frac{\Gamma(-
\alpha' s) \Gamma(- \alpha' t)}{\Gamma(1- \alpha' s - \alpha' t)}
K(\zeta_1, k_1; \zeta_2, k_2; \zeta_3, k_3; \zeta_4, k_4) \ ,
\label{A4fermionic}
\end{eqnarray}
where $K$ is the $4$-point kinematic factor (which contain no poles) given in formula (\ref{Ks}) of Appendix \ref{Gauge boson}.\\
The $s$ and $t$ appearing in (\ref{A4fermionic}) are part of the Mandelstam variables:
\begin{eqnarray}
s = - (k_1 +k_2)^2 \ , \ \ \ t = - (k_1+k_4)^2 \ , \ \ \ u = -(k_1+k_3)^2 \ .
\label{Mandelstam}
\end{eqnarray}
These variables satisfy the condition
\begin{eqnarray}
s + t + u = 0 \ .
\label{condition}
\end{eqnarray}
Using momentum conservation, the physical state and the mass shell conditions ($\zeta_j \cdot k_j = 0$ and $k_j^2=0$, respectively), it is not difficult to see that the $3$ and $4$-point subamplitudes that we have presented in this subsection satisfy (on-shell) gauge invariance, cyclic symmetry and world-sheet parity (that in ref.\cite{Barreiro1} we have called `twisting' symmetry).

\subsection{Explicit structure of the low energy effective lagrangian up to ${\alpha'}^3$ order}

\label{Explicit structure}

In this subsection we will concentrate on the bosonic terms of the OSLEEL.
Here we will not give the details of how to construct a complete list of independent terms at each order in $\alpha'$ in ${\cal L}_{\mbox{eff}}$. It is well known that the procedure to do so involves using integration by parts, the Bianchi identity, (\ref{Bianchi}), and the $[D,D] \cdot  = [F, \cdot ]$ identity, (\ref{group4}). In the literature this has been done up to ${\alpha'}^4$ terms (see \cite{Tseytlin1},\cite{ Koerber1}, \cite{DeRoo1} and \cite{Koerber2}, for example). We will quote a reference whenever we need this list of terms at some order in $\alpha'$.\\
It is important to remark that the explicit structure of the terms of the low energy effective lagrangian will not just be the one that one arrives to following the procedure mentioned up to here since this lagrangian needs only to reproduce $on-shell$ scattering amplitudes. It turns out that some terms of the lagragian are not sensible to S-matrix calculations, or equivalently, their coefficients will not remain unchanged under field redefinitions, so they are of no importance for the low energy effective lagrangian. Those terms will be discarded after being identified.

\subsubsection{Low energy effective lagrangian up to ${\alpha'}^2$ terms}

\label{Low energy1}

The calculation on the nonabelian Born-Infeld lagrangian up to ${\alpha'}^2$ terms has been done in many places in the literature. We will follow section 7 of Tseytlin's paper \cite{Tseytlin1}, where it was found that one possibility of writing down the lagrangian as a sum of independent terms is the following:
\begin{eqnarray}
{\cal L}_{\mbox{eff}} =  \frac{1}{g^2} \mbox{tr} \biggr[   &-&\frac{1}{4} F^{\mu\nu}F_{\mu\nu}
+(2\alpha')^ 1 \Big( a_{1}F_{\mu}^{\ \lambda}F_{\lambda}^{\ \nu}F_{\nu}^{\ \mu}
+a_{2}D^{\lambda}F_{\lambda}^{\ \mu}D^{\rho}F_{\rho\mu}  \Big)+ \nonumber \\
&&(2\alpha')^{2}\Big( a_{3}F^{\mu\lambda}F^{\nu}_{\ \lambda}F_{\mu}^{\ \rho}F_{\nu\rho} + +a_{4}F^{\mu}_{\ \lambda}F_{\nu}^{\ \lambda}F^{\nu\rho}F_{\mu\rho}
+a_{5}F^{\mu\nu}F_{\mu\nu}F^{\lambda\rho}F_{\lambda\rho}+ \nonumber \\
&&a_{6}F^{\mu\nu}F^{\lambda\rho}F_{\mu\nu}F_{\lambda\rho}
+a_{7}F^{\mu\nu}D^{\lambda}F_{\mu\nu}D^{\rho}F_{\rho\lambda}+
+a_{8}D^{\lambda}F_{\lambda\mu}D^{\rho}F_{\rho\nu}F^{\mu\nu}+ \nonumber \\
&&a_{9}D^{\rho}D^{\lambda}F_{\lambda}^{\mu}D_{\rho}D^{\sigma}F_{\sigma\mu}\Big)+O({(2\alpha')}^{3}) \ \biggr] \ .
\label{L12}
\end{eqnarray}
\noindent It was seen in \cite{Tseytlin1} that, after examining the possibility of field redefinitions, the coefficients $\{ a_2, a_7, a_8, a_9 \}$ remain arbitrary and thus are not sensible to any S-matrix calculation, so they may be chosen to be zero, leading to
\begin{eqnarray}
{\cal L}_{\mbox{eff}} =  \frac{1}{g^2} \mbox{tr} &\biggr[ &  -\frac{1}{4} F^{\mu\nu}F_{\mu\nu}
+(2 \alpha')^1 a_{1}F_{\mu}^{\ \lambda}F_{\lambda}^{\ \nu}F_{\nu}^{\ \mu}+ \nonumber \\
&&(2\alpha')^{2}\Big( a_{3}F^{\mu\lambda}F^{\nu}_{\ \lambda}F_{\mu}^{\ \rho}F_{\nu\rho}
 +a_{4}F^{\mu}_{\ \lambda}F_{\nu}^{\ \lambda}F^{\nu\rho}F_{\mu\rho}   + \nonumber\\
&& a_{5}F^{\mu\nu}F_{\mu\nu}F^{\lambda\rho}F_{\lambda\rho}
+a_{6}F^{\mu\nu}F^{\lambda\rho}F_{\mu\nu}F_{\lambda\rho} \Big) + O({(2\alpha')}^{3}) \ \biggr] \ .
\label{L12simpler}
\end{eqnarray}

\subsubsection{Low energy effective lagrangian at ${\alpha'}^3$ order}

\label{Low energy2}

In \cite{Koerber1}, for the ${\alpha'}^3$ terms in (\ref{L12simpler}) it was found a 36-dimensional basis (containing 6 $F^5$ terms, 24 $D^2 F^4$ terms, 5 $D^4 F^3$ terms and 1 $D^6 F^2$ term). After discarding the terms which are sensible to field redefinitions, in \cite{Koerber1} it was seen that that the remaining lagrangian has the following 13 terms:
\begin{eqnarray}
{\cal L_{\mbox{eff}}}^{(3)}\hspace{-2pt}&=&\hspace{-2pt}\frac{(2 \alpha')^3}{g^2} \hspace{-1.5pt} \mbox{tr} \Big[\hspace{-0.3pt} a_{10} \hspace{-0.3pt} F_{\mu}^{\ \nu}F_{\nu}^{\ \lambda}F_{\lambda}^{\ \rho}F_{\rho}^{\ \sigma}F_{\sigma}^{\ \mu} \hspace{-1pt}+\hspace{-1pt} a_{11}\hspace{-2pt} \ F_{\mu}^{\ \nu}F_{\nu}^{\ \lambda}F_{\lambda}^{\ \rho}F_{\sigma}^{\ \mu}F_{\rho}^{\ \sigma} \hspace{-1pt}+\hspace{-1pt}
a_{12}\hspace{-2pt} \ F_{\mu}^{\ \nu}F_{\nu}^{\ \lambda}F_{\sigma}^{\ \mu}F_{\lambda}^{\ \rho}F_{\rho}^{\ \sigma} + \nonumber \\
& &\hphantom{({2 \alpha'})^3 \mbox{tr} \biggl[ } \hspace{-2pt} a_{13}\hspace{-2pt} \ F_{\mu}^{\ \nu}F_{\rho}^{\ \sigma}F_{\nu}^{\ \lambda}F_{\sigma}^{\ \mu}F_{\lambda}^{\ \rho} \hspace{-1pt}+\hspace{-1pt} a_{14}\hspace{-2pt} \ F_{\mu}^{\ \nu}F_{\nu}^{\ \lambda}F_{\lambda}^{\ \mu}F_{\rho}^{\ \sigma}F_{\sigma}^{\ \rho} \hspace{-1pt}+\hspace{-1pt}
a_{15}\hspace{-2pt} \ F_{\mu}^{\ \nu}F_{\nu}^{\ \lambda}F_{\rho}^{\ \sigma}F_{\lambda}^{\ \mu}F_{\sigma}^{\ \rho}\hspace{-1pt} + \nonumber \\
& &\hphantom{({2 \alpha'})^3 \mbox{tr} \biggl[ }  a_{16} ( D_{\mu}F_{\nu}^{\ \lambda} ) ( D^{\mu}F_{\lambda}^{\ \rho} ) F_{\sigma}^{\ \nu}F_{\rho}^{\ \sigma} \hspace{-0.5pt}+\hspace{-0.5pt}
a_{17} (D_{\mu}F_{\nu}^{\ \lambda})F_{\sigma}^{\ \nu}(D^{\mu}F_{\lambda}^{\ \rho})F_{\rho}^{\ \sigma} + \nonumber \\
& &\hphantom{({2 \alpha'})^3 \mbox{tr} \biggl[ }  a_{18} (D_{\mu}F_{\nu}^{\ \lambda})(D^{\mu}F_{\lambda}^{\ \nu})F_{\rho}^{\ \sigma}F_{\sigma}^{\ \rho} \hspace{-0.5pt}+\hspace{-0.5pt}
a_{19} (D_{\mu}F_{\nu}^{\ \lambda})F_{\rho}^{\ \sigma}(D^{\mu}F_{\lambda}^{\ \nu})F_{\sigma}^{\ \rho} + \nonumber \\
& &\hphantom{({2 \alpha'})^3 \mbox{tr} \biggl[ }  a_{20} (D_{\sigma}F_{\mu}^{\ \nu})F_{\lambda}^{\ \rho}(D^{\mu}F_{\nu}^{\ \lambda})F_{\rho}^{\ \sigma} \hspace{-0.5pt}+\hspace{-0.5pt}
	a_{21} F_{\mu}^{\ \nu}(D^{\mu}F_{\nu}^{\ \lambda})F_{\rho}^{\ \sigma}(D_{\sigma}F_{\lambda}^{\ \rho} ) + \nonumber \\
& &\hphantom{({2 \alpha'})^3 \mbox{tr} \biggl[ }  a_{22}
F_{\mu}^{\ \nu}(D^{\mu}F_{\lambda}^{\ \rho})(D_{\sigma}F_{\nu}^{\ \lambda}
)F_{\rho}^{\ \sigma} \ \Big] \
. \nonumber \\
\label{L3}
\end{eqnarray}

\subsection{Basics of the conventional S-matrix approach to the low energy effective lagrangian}

\label{S-matrix}

The conventional S-matrix method consists in determining the coefficients of the low energy effective lagragian using the known expression of the on-shell scattering amplitudes from String Theory. More specifically, in the case of the bosonic terms, in order to determine the ${\alpha'}^{N}$ order terms in  ${\cal L}_{\mbox{eff}}$ it is necessary to know the previous ${\alpha'}^k$ order terms of it ($k=1, \ldots, N-1$)\footnote{In the case of the open superstring it is
only necessary to know the previous ${\alpha'}^k$ order terms, with $k=1, \ldots, N-2$.} and it is also necessary to know the $(N+2)$-point gauge boson amplitude (where $N \ge 1$) from Open String theory, expanded at that order in $\alpha'$.  For example, to determine the $(2 \alpha') F^3$ terms it is necessary to know the $3$-point amplitude, to determine the  $(2\alpha')^2 F^4$ terms it is necessary to know the $4$-point amplitude, and so on. What one does is to compute the $(N+2)$-point gauge boson subamplitude from the ${\alpha'}^{N}$ order terms of  ${\cal L}_{\mbox{eff}}$ and then compares this expression (which is a linear function of the unknown coefficients $a_j$) with the corresponding one coming from String Theory at that $\alpha'$ order. This determines $uniquely$ those coefficients. For example, in the case of the $3$-point subamplitude that comes from  ${\cal L}_{\mbox{eff}}$ in (\ref{L12simpler}), using Feynman rules it is easy to arrive at the following expression for it\footnote{Along this work $A_{YM}(1, \ldots, N)$ will denote the tree level Yang-Mills $N$-point subamplitude.}:
\begin{eqnarray}
A(1,2,3) & = & A_{YM}(1,2,3) + 6\ i \ a_1 \ g \ (2 \alpha')^1 \ (\zeta_1 \cdot k_2) (\zeta_2 \cdot k_3) (\zeta_3 \cdot k_1) \ .
\label{A3}
\end{eqnarray}
Comparing this expression with the one for $A(1,2,3)$, given in (\ref{A3fermionic}), respectively, we find that $a_1=0$ for the supersymmetric open string. In this case there is only one equation for $a_1$. \\
It is known that the analogue procedure in the case of the $4$-point subamplitude, leads to \cite{Tseytlin2}
\begin{eqnarray}
a_3 = \pi^2/12 \ , \ \ a_4 = \pi^2/24 \ , \ \ a_5 = -\pi^2/48 \ , \ \ a_6 = -\pi^2/96 \ .
\label{table1}
\end{eqnarray}
An important fact, about the determination of the previous coefficients, is that the linear system of equations for them is \underline{overdetermined} (consistently). This means that it is not necessary to know the complete expression of the amplitude in order to find the coefficients. For example, in the case of the $4$-point amplitude it is sufficient to compare the $(\zeta \cdot \zeta)^2$ terms of (\ref{A4fermionic})  with the corresponding ones of $A(1,2,3,4)$, at $\alpha'$ and ${\alpha'}^2$ order \cite{Tseytlin2}. What happens is that the terms of the amplitude that were not considered to find the coefficients $a_j$ can all be determined from the first ones by demanding (on-shell) gauge invariance and cyclic symmetry.\\
For $N \ge 5$, at first sight, there arise two complications in the S-matrix computations of Open Superstring Theory:
\begin{enumerate}
\item The number of terms that appear in the $N$-point subamplitude grows considerably.
\item The determination of the numerical coefficients of each kinematical term of the subamplitude is not straight forward (as in the $N=4$ case).
\end{enumerate}
For example, in the case of $N=5$ direct application of formula (\ref{ANfermionic}) leads to an expression which has more than 140 terms, of the following form\footnote{See eq. (5.29) of \cite{Brandt1} for the complete detailed formula.}:
\begin{eqnarray}
A(1,2,3,4,5)  =  2 g^3 (2  \alpha')^2 \Big\{\hspace{-5pt}&& L_3 (\zeta_1 \cdot \zeta_2)(\zeta_3 \cdot \zeta_4)(\zeta_5 \cdot k_2)(k_1 \cdot k_3) \nonumber \\
&& \big( (\zeta \cdot \zeta)^2 (\zeta \cdot k)(k \cdot k) \ \mbox{terms} \big) + \nonumber \\
&& K_2 (\zeta_1 \cdot \zeta_4)(\zeta_5 \cdot k_2)(\zeta_2 \cdot k_1)(\zeta_3 \cdot k_4) \nonumber \\
&& \big( (\zeta \cdot \zeta) (\zeta \cdot k)^3  \big) \Big\} .
\label{A12345-1}
\end{eqnarray}
In this last formula $L_3$ and $K_2$ are momentum dependent factors (which also depend on $\alpha'$) given by double integrals:
\begin{eqnarray}
\biggl\{ \begin{array}{c}
         L_3 \\
         K_2
         \end{array} \biggr\}
\hspace{-3pt}=\hspace{-5pt} \int_0^1 \hspace{-3pt} dx_3 \hspace{-3pt} \int _0^{x_3}\hspace{-3pt} dx_2 \hspace{-0.5pt} x_2^{2 \alpha' \alpha_{12}} (1\hspace{-3pt}-\hspace{-3pt}x_2)^{2 \alpha' \alpha_{24}} x_3^{2 \alpha' \alpha_{13}} (1\hspace{-3pt}-\hspace{-3pt}x_3)^{2 \alpha' \alpha_{34}} (x_3\hspace{-3pt}-\hspace{-3pt}x_2)^{2 \alpha' \alpha_{23}}
\biggl\{ \begin{array}{c}
         \frac{1}{x_2 x_3 (1-x_3)} \\
         \frac{1}{x_2 (1-x_3)}
         \end{array} \biggr\}  ,  \nonumber \\
\label{L3K2}
\end{eqnarray}
where $\alpha_{ij}=k_i \cdot k_j$. They can be calculated in terms of Beta and Hypergeometric functions. Although not immediately, the coefficients of the first terms of their $\alpha'$ expansion can be obtained\footnote{These coefficients can be calculated, for example, using techniques of Harmonic Polylogarithms \cite{Remiddi1} or Harmonic Sums \cite{Vermaseren1}, which are nowadays perfectly understood.}, for example \cite{Brandt1}
\begin{eqnarray}
K_2 & = & \frac{1}{(2 \alpha')^2} \left\{ \frac{1}{\alpha_{12} \
\alpha_{34}} \right\} - \zeta(2) \left\{\frac{
\alpha_{51} \ \alpha_{12} - \alpha_{12} \ \alpha_{34} +
\alpha_{34} \ \alpha_{45}}{\alpha_{12} \ \alpha_{34}} \right\} +
\nonumber
\\&& \zeta(3) \ (2 \alpha') \left\{  \frac{\alpha_{12}^2 \hspace{-0.4pt} \alpha_{51}
 \hspace{-3pt}-\hspace{-3pt} \alpha_{34}^2 \hspace{-0.4pt} \alpha_{12} \hspace{-3pt}+\hspace{-3pt}
\alpha_{45}^2 \hspace{-0.4pt} \alpha_{34} \hspace{-3pt}+\hspace{-3pt} \alpha_{51}^2 \hspace{-0.4pt} \alpha_{12} \hspace{-3pt}-\hspace{-3pt}
\alpha_{12}^2 \hspace{-0.4pt} \alpha_{34} \hspace{-3pt}+\hspace{-3pt} \alpha_{34}^2 \hspace{-0.4pt} \alpha_{45} \hspace{-3pt}-\hspace{-3pt} 2
\alpha_{12} \hspace{-0.4pt}  \alpha_{23} \hspace{-0.4pt}  \alpha_{34} }{\alpha_{12}
\alpha_{34}} \right\} \
\hspace{-3pt}+\hspace{-3pt} \nonumber\\
&& {\cal O}((2\alpha')^2) \,.
\label{K2}
\end{eqnarray}
The complication mentioned above, in item 1, has been recently circumvented in ref.\cite{Mafra1} by finding a  general compact formula for the open superstring $N$-point subamplitude: it can be shortly written in terms of a basis of $(N-3)!$ Yang-Mills subamplitudes, each of them being multiplied by an identified $(N-3)$-dimensional multiple integral\footnote{In fact, the result of ref. \cite{Mafra1} is much more complete, in the sense that the whole $N$-point subamplitude (the one that includes gauge bosons and gauginos) has been calculated in terms of Super Yang-Mills $N$-point subamplitudes, using a Pure Spinor formalism.}. So this avoids enormously dealing with long expressions like the one in (\ref{A12345-1}), but still, in order to obtain the $(2 \alpha')^{N-2} F^N$ terms of the low energy effective lagrangian, it is necessary to compute the ${\alpha'}^{N-2}$ coefficients of the mentioned $(N-3)$-dimensional multiple integrals (like the ones in (\ref{L3K2}) and (\ref{integral principal})).\\
For example, besides the case of $K_2$ (mentioned in (\ref{L3K2}) and (\ref{K2})), in the case of the $6$-point amplitude, one of the many triple integrals that appears is the following \cite{Stieberger2, Barreiro2} :
\begin{eqnarray}
I_3 &=&\int_{0}^{1}\hspace{-3pt}dx_4\int_{0}^{x_4}\hspace{-3pt}dx_3\int_{0}^{x_3}\hspace{-3pt} dx_2 \ {x_2}^{2 \alpha' \alpha_{12}} \ {x_3}^{2 \alpha'  \alpha_{13}} \ {x_4}^{2 \alpha' \alpha _{14}-1}  \ {(x_3\hspace{-3pt}-\hspace{-3pt}x_2)}^{2 \alpha'  \alpha_{23} - 1} \ {(x_4\hspace{-3pt}-\hspace{-3pt}x_2)}^{2 \alpha'  \alpha_{24}} \cdot \nonumber \\
&& \hphantom{ \int_{0}^{1}dx_4\int_{0}^{x_4}dx_3\int_{0}^{x_3}dx_2 \ }  {(x_4\hspace{-3pt}-\hspace{-3pt}x_3)}^{2 \alpha'  \alpha_{34}-1}   {(1\hspace{-3pt}-\hspace{-3pt}x_2)}^{2 \alpha'  \alpha_{25}-1} \ {(1\hspace{-3pt}-\hspace{-3pt}x_3)}^{2 \alpha'  \alpha_{35}}{(1\hspace{-3pt}-\hspace{-3pt}x_4)}^{2 \alpha'  \alpha_{45}} \ , \nonumber \\
\label{integral principal}
\end{eqnarray}
which has an $\alpha'$ expansion which begins like \cite{Hemily1}\footnote{In the literature, the $\alpha'$ expansions of tree level 6-point integrals seem to have first appeared in ref. \cite{Stieberger2}. On this reference a detailed study of the open superstring $6$-point amplitude was done. It was also explained there how to compute the coefficients of the kinematical terms using Euler sums, but due to the lengthness of the formulas, not all of the $\alpha'$ expansions were explicitly given. The result we have cited for eq. (\ref{I3}), namely ref. \cite{Hemily1}, has calculated this expansion independently of the calculations of ref. \cite{Stieberger2}, using Harmonic Sums techniques \cite{Vermaseren1}.}
\begin{eqnarray}
I_3 &=& \frac{1}{(2\alpha')^3}\biggl[ \biggl( \frac{1}{\alpha_{23}\alpha_{16}t_{234}}+\frac{1}{\alpha_{34}\alpha_{56}t_{234}}\biggr) +\biggl( \frac{1}{\alpha_{34}\alpha_{16}t_{234}}+\frac{1}{\alpha_{23}\alpha_{56}t_{234}}\biggr) \biggr] + \nonumber \\
&& \frac{\zeta(2)}{(2\alpha')^1}\biggl[-\biggl(\frac{\alpha_{16}}{\alpha_{23}t_{234}} \hspace{-3pt}+\hspace{-3pt}\frac{\alpha_{56}}{t_{234}\alpha_{34}}\biggr)
\hspace{-3pt}+\hspace{-3pt} \biggl(\frac{1}{\alpha_{56}}\hspace{-3pt}+\hspace{-3pt}\frac{1}{\alpha_{16}}\biggr)\hspace{-3pt}-\hspace{-3pt} \biggl(\frac{\alpha_{23}}{t_{234}\alpha_{56}}\hspace{-3pt}+\hspace{-3pt}\frac{\alpha_{34}}{t_{234}\alpha_{16}}\biggr) \hspace{-3pt}+\hspace{-3pt}\biggl(\frac{1}{\alpha_{23}}\hspace{-3pt}+\hspace{-3pt}\frac{1}{\alpha_{34}}\biggr) \hspace{-3pt}-\hspace{-3pt} \nonumber \\
&& \hphantom{  \frac{1}{(2\alpha')}\biggl[  }   \biggl(\frac{\alpha_{23}}{t_{234}\alpha_{16}}\hspace{-3pt}+\hspace{-3pt}\frac{\alpha_{34}}{t_{234}\alpha_{56}}\biggr) \hspace{-3pt}-\hspace{-3pt}\biggl(\frac{\alpha_{12}}{\alpha_{56}\alpha_{34}} \hspace{-3pt}+\hspace{-3pt}\frac{\alpha_{45}}{\alpha_{23}\alpha_{16}}\biggr)\hspace{-3pt}-\hspace{-3pt} \biggl(\frac{\alpha_{56}}{\alpha_{23}t_{234}}\hspace{-3pt}+\hspace{-3pt}\frac{\alpha_{16}}{t_{234}\alpha_{34}}\biggr) \hspace{-3pt}-\hspace{-3pt} \nonumber \\
&&  \hphantom{  \frac{1}{(2\alpha')}\biggl[  }    \biggl(\frac{t_{345}}{\alpha_{34}\alpha_{16}}\hspace{-3pt}+\hspace{-3pt}\frac{t_{123}}{\alpha_{23}\alpha_{56}}\biggr)\biggr] + O((2\alpha')^0) \ ,
\label{I3}
\end{eqnarray}
where, besides $\alpha_{ij}=k_i \cdot k_j$, in (\ref{I3}) we are calling $t_{ijk}= \alpha_{ij}+\alpha_{ik}+\alpha_{kj}$ \cite{Stieberger2}.\\
So, at the end, in order to calculate the coefficients of the open string low energy effective lagrangian at a given order in $\alpha'$, the main difficulty that nowadays exists is the one mentioned above, in item 2, namely, finding the explicit $\alpha'$ expansion of certain $(N-3)$-multiple integrals (for $N \ge 7$).

\noindent So this is the basics of the conventional S-matrix method. Although here we only mentioned the scattering amplitudes of gauge bosons, in the case of Open Superstring Theory it applies exactly in the same way to obtain the fermionic terms of the OSLEEL, by considering the scattering amplitudes of bosons and fermions. We have called it `conventional' in order to distinguish it from the `revisited' one, that we will present in this work. \\
What we would like to remark at this point is that, independently of working in Bosonic or in Supersymmetric Open String Theory\footnote{The $N$-point amplitude for gauge bosons in the case of Bosonic open String Theory has a known expression, similar to the one we have reviewed in eq.(\ref{ANfermionic}).}, as long as we are considering only interactions of gauge bosons, the conventional S-matrix approach to the low energy effective lagrangian makes no difference between the calculations done with done with $A(1, \ldots, N)$ in the bosonic or in the supersymmetric theory: in both cases one has to deal with $\alpha'$ expansions of $(N-3)$-multiple integrals and with kinematic expressions and one has to match these amplitudes with the ones that come from \underline{the same} general low energy effective lagrangian, that is, the one that is given by eqs. (\ref{L12simpler}) and (\ref{L3}), and higher order $\alpha'$ terms. Apparently, there is no peculiarity (besides dealing with extra Grassmann variables) in computing gauge boson interaction terms when one deals with a supersymmetric theory.

\section{Revisiting the S-matrix approach to the low energy effective lagrangian}

\label{A new}

From the list of methods to arrive to the nonabelian low energy effective lagrangian that were mentioned in the Introduction (section \ref{Introduction}), for Open Superstring Theory, only the S-matrix approach and the one that deals with BPS configurations (due to Koerber and Sevrin) are capable of finding the bosonic terms without having to deal directly with the fermionic ones. In this section we will present our revisited S-matrix approach. \\
The main observation of our revisited S-matrix approach is that, even if we are only dealing with pure gauge boson interactions, \underline{there is difference} in the calculations between the S-matrix approach (at tree level) to ${\cal L}_{\rm{eff}}$ in the bosonic and the supersymmetric theory of open strings. We will see that in the case of the supersymmetric theory of open strings the $a_j$ coefficients of ${\cal L}_{\rm{eff}}$ satisfy constraints that come from the kinematical structure of the gauge boson $N$-point amplitude, $A(1, \ldots, N)$.\\
The constraints for the  $a_j$ coefficients of ${\cal L}_{\rm{eff}}$ that we will refer to in the next subsection are similar, if not the same, to the ones found in \cite{Koerber1, Koerber2} by the method of BPS configurations.\\
Due to the fact that the equations for the $a_j$ coefficients are overdetermined (as we mentioned in the previous subsection), the constraints that we have just mentioned will diminish the number of unknowns, at a given $\alpha'$ order, and it will not be necessary to calculate $\alpha'$ expansions of multiple integrals like the ones in (\ref{L3K2}) and (\ref{integral principal}) (or even more complicated ones which appear in higher $N$-boson subamplitudes, with $N \ge 7$) to determine the value of the $a_j$'s.  Our method seems to need only $4$-point subamplitudes to fix the $a_j$ coefficients. At  least we have confirmed this up to ${\alpha'}^4$ terms, just like in \cite{Koerber1, Koerber2}.

\subsection{The basic idea of the method: An important constraint arises in the gauge boson amplitudes of Open Superstring Theory}

\label{An important}

If we consider the $N$-point subamplitude of gauge bosons in the open superstring (see eq.(\ref{ANfermionic})), it is easy to see that the $(\zeta \cdot k)^N$ terms will only come out from integrating on the $x_i$ variables a term of the following type:
\begin{eqnarray}
T & \sim &  \int d \theta_1 \ldots d \theta_{N-2}  \int d \phi_1 \ldots d \phi_N \ P_N(x, \theta) \ \Big(   \sum_{i \neq j}^N
\frac{  (2 \alpha')^1  (\theta_j-\theta_i) \phi_j (\zeta_j \cdot k_i)  }{x_j-x_i}    \Big)^N \ .
\label{T}
\end{eqnarray}
Here $P_N(x, \theta)$ corresponds to the $\theta$ expansion of the product of terms $( x_q - x_p - \theta_q \theta_p )^{2 \alpha' k_p \cdot k_q}$ in (\ref{ANfermionic}), where $\theta_{N-1}=\theta_N=0$\footnote{See the second footnote on page 5.}. \\
From the point of view of the $\phi_j$ Grassmann variables in (\ref{T}), after expanding the $()^N$ term, it is easy to see that the coefficient of $\phi_1 \phi_2 \ldots \phi_N$ (which is the only nonzero product of $N$ $\phi_j$ variables) only contains products of $N$ $\theta_i$'s. Since  $\theta_{N-1}=\theta_N=0$ this coefficient is always zero and, therefore, $T=0$.\\
So our main conclusion is that
\begin{eqnarray}
\mbox{$A(1, \ldots, N)$  does not contain $(\zeta \cdot k)^N$ terms  } \ ,
\label{constraint}
\end{eqnarray}
where $A(1, \ldots, N)$ is the $N$-point gauge boson amplitude in Open Superstring Theory.\\
It is easy to check that the $3$ and $4$-point amplitudes in Bosonic Open String Theory do not satisfy the constraint in (\ref{constraint})\footnote{See, for example, ref. \cite{Tseytlin2} for the 3-point amplitude and ref. \cite{Kawai1} for the 4-point amplitude.}, respectively, and from the general formula for the $N$-point gauge boson amplitude in this theory \cite{Schwarz1} we do not expect that the $(\zeta \cdot k)^N$ terms to cancel among themselves, so we really expect that the constraint in (\ref{constraint}) does not happen in the case of Bosonic Open String Theory and that it is only a peculiarity of Open Superstring Theory\footnote{The fact that a cancellation of the $(\zeta \cdot k)^N$ terms, among themselves, might happen, is already known to occur in the case of the $4$-point amplitude. In Appendix \ref{Gauge boson}, in equations (\ref{Ks}) and (\ref{manifest2}) we have written the kinematic factor of this amplitude in two different, but on-shell equivalent, ways: the first one with no $(\zeta \cdot k)^4$ terms, but without manifest gauge invariance, and the second one containing $(\zeta \cdot k)^4$ terms and being manifestly gauge invariant.}.\\
It is this constraint that makes all the difference between the `conventional' and the `revisited' S-matrix approach to the low energy effective lagrangian in Open String Theory.  We will use it to find relations between the different coefficients of the bosonic terms presented in eqs. (\ref{L12simpler}) and (\ref{L3}), at each order in $\alpha'$. We will do this in section \ref{Applying} and we will then determine the ${\alpha'}^4$ terms of the low energy effective lagrangian in section \ref{Low}.\\

\subsection{How the method works}

\label{How}

Let us consider the bosonic part of the low energy effective lagrangian ${\cal L}_{\mbox{eff}}$. At order ${\alpha'}^p$ ($p \ge 2$) its general term consists of $(p-1)$ subterms $D^{2i-2}F^{p+3-i}$ ($i=1, \ldots, p-1$) :
\begin{eqnarray}
{\cal L}_{\mbox{eff}}^{(p)} & = & \frac{1}{g^2} (2 \alpha')^p \biggl[ F^{p+2} + D^2 F^{p+1} + D^4 F^p + \ldots + D^{2p-4} F^4 \biggr] \ .
\label{Lk}
\end{eqnarray}
As explicited in some cases in subsection \ref{Explicit structure}, each subterm $D^{2i-2}F^{p+3-i}$ really means a linear combination of independent terms of that type (with, up to now, unknown coefficients). In order to determine ${\cal L}_{\mbox{eff}}^{(p)}$ our `revisited' S-matrix method works in two steps:\\

\begin{itemize}
\item[I.] \emph{Reduction from the general basis to the constrained basis.} In this step is where we demand the constraint (\ref{constraint}) in the $N$-point subamplitude calculated from ${\cal L}_{\mbox{eff}}^{(p)}$ at order ${\alpha'}^p$ \footnote{From $p=4$ onwards, the method requires also the knowledge of  ${\cal L}_{\mbox{eff}}^{(2)}, \ldots,  {\cal L}_{\mbox{eff}}^{(p-2)}$, in order to compute the ${\alpha'}^p$ contribution on the $N$-point subamplitude..}, where $N=4, \ldots, (p+2)$. This will lead to strong constraints between the unknown coefficients of the lagrangian. These constraints consist on a system of $(N-2)^N$ linear equations for the $a_j$ coefficients\footnote{After using momentum conservation to eliminate, say $k_N$ in terms of the other $k_i$'s, and demanding the physical state condition, there are only left $(N-2)^N$ different $(\zeta \cdot k)^N = (\zeta_1 \cdot k_{i_1}) \ldots (\zeta_N \cdot k_{i_N})$ terms.}. For increasing $N$ (starting from $N \ge 6$, for example) this number of equations grows extremely rapidly, but it is guaranteed that it always has nonzero solutions for at least some of the $a_j$ coefficients because, in particular, the $N$-point amplitudes coming from the OSLEEL obey it and it is known that in this lagrangian the ${\alpha'}^p D^{2p-4} F^4$ terms are non zero\footnote{It is well known that the $\alpha'$ expansion of the $4$-point amplitude has non zero coefficients from $p=2$ onwards.}. The previous lines imply that the equations coming from demanding absence of $(\zeta \cdot k)^N$ terms in the $N$-point subamplitude should not be all linearly independent. In fact this is expected to happen because the full subamplitude satisfies, among various properties, cyclic invariance and (on-shell) gauge invariance, and this relates the coefficients of its kinematical terms (in particular implying that the coefficients of the $(\zeta \cdot k)^N$ terms are not all independent; that is, if some of the coefficients of these terms are zero, then cyclic and gauge invariance can be used to derive that necessarily some other coefficients of these type of kinematical terms are also zero).
    Anyway, the solution to these constraints is such that at the end only a few of the $a_j$ coefficients will still remain unknown, according  to the following table\footnote{There are some observations with respect to the table in (\ref{table2}):1. \  We have included the case of $p=1$, not considered in eq. (\ref{Lk}), which simply states that there are no $F^3$ terms at order ${\alpha'}^1$ in the constrained case.
    2. \ The informations about ${\alpha'}^4$ order have been taken from \cite{Koerber2}, which we expect to agree completely with our results.
    3. \ At first glance, the fact that for $p=4$ the constrained basis has dimension $0$ might seem surprising since this number is $less$ than the corresponding one for $p=2$ and $p=3$. In section \ref{Low}, in the second paragraph after eq.(\ref{LD4F4}) we explain the reason for this: it has to do with the fact that from $p=4$ onwards, the constraint in eq. (\ref{constraint}) makes the $a_j$ coefficients of ${\cal L}_{\mbox{eff}}^{(p)}$ dependent on the ones from   ${\cal L}_{\mbox{eff}}^{(2)}, \ldots,  {\cal L}_{\mbox{eff}}^{(p-2)}$, as mentioned in footnote $20$, on page 11.}:
    \begin{eqnarray}
    \begin{tabular}{|c|c|c|}
    \hline
    $p$ & Dimension of the general basis & Dimension of the constrained basis\\
      & at order ${\alpha'}^p$              & at order  ${\alpha'}^p$ \\
      \hline
      1 & 1 & 0 \\
      2 & 4 & 1 \\
      3 & 13 & 1 \\
      4 & 96 & 0 \\
      \vdots & \vdots & \vdots
      \label{table2} \\
      \hline
      \end{tabular}
      \end{eqnarray}
      %\vspace{0.5cm}
      \noindent On the previous table we understand by `dimension' the number of independent terms, at a given ${\alpha'}^p$ order, whose coefficients are not sensible to field redefinitions. The column saying `general basis' specifies the number of coefficients that the `conventional' S-matrix approach would require to determine at that order, by means of a $(p+2)$-point open string amplitude\footnote{The dimension of the general basis in the case of $p=1, 2, 3$ is precisely the number of undetermined coefficients found in subsections \ref{Low energy1} amd \ref{Low energy2}.}. The column saying `constrained basis' specifies the number of coefficients that the `revisited' S-matrix method requires to determine.

      Notice that on this step we are indeed using the open superstring $N$-point subamplitude, given in (\ref{ANfermionic}), where $N=4, \ldots, (p+2)$, as the `conventional' S-matrix approach does. The important detail is that, due to the redundancy of information on that formula we do not need to explicitly compute the complete expression for the scattering subamplitude: we only use the part of it which is convenient to us, that is, the absence of the $(\zeta \cdot k)^N$ terms. We do not expect that demanding the constraint in (\ref{constraint}) for $N > p+2$ will lead to new (linearly independent) conditions for the $a_j$ coefficients at ${\alpha'}^p$ order\footnote{This is similar to the fact that, after knowing the explicit expression of the open superstring $(p+2)$-point subamplitude at ${\alpha'}^p$ order, any information obtained from a higher $N$-point amplitude at that $\alpha'$ order is redundant.}.

      Having accepted that the kinematical constraints in (\ref{constraint}) lead to non trivial solutions for the $a_j$ coefficients it is natural to raise the question of how many undetermined coefficients we are left with, or stated in another way, what is the dimension of the constrained basis. In the table in (\ref{table2}) we see that for low values of $N$ this number is $1$ or $0$, but at this moment we do not have a clear answer about this dimension for higher orders of $\alpha'$.
    \item[II.] \emph{Determination of the coefficients of the constrained basis.} After $step \ I$ the remaining free coefficients in ${\cal L}_{\mbox{eff}}^{(p)}$ would have to be determined using the explicit expression of a lower open superstring $N$-point subamplitude ($N<p+2$) at ${\alpha'}^p$ order. For example, in the next two sections we will see that knowledge of the open superstring $4$-point subamplitude is enough to find not only the ${\alpha'}^2$ (as it happens with the conventional S-matrix approach) but also the ${\alpha'}^3$ and the ${\alpha'}^4$ terms of the low energy effective lagrangian\footnote{In \cite{Koerber2} it was also seen that the $4$-point subamplitude is enough to find the ${\alpha'}^4$ terms, once the BPS constraints in the bosonic terms are taken into account.}.\\
\end{itemize}

\section{Applying the revisited S-matrix approach to obtain ${\cal L}_{\mbox{eff}}$ up to ${\alpha'}^3$ order}

\label{Applying}

\subsection{Low energy effective lagrangian up to ${\alpha'}^2$ order}

\label{Low energy}

In this subsection we will reproduce, once again, the results mentioned in table (\ref{table1}) for the lagrangian in (\ref{L12simpler}) (in the case of Open Superstring Theory), but this time strictly following the two steps of the `revisited' method, mentioned in subsection \ref{How}. \\
We start considering the $3$-point subamplitude. We already saw in subsection \ref{S-matrix} that the absence of $(\zeta \cdot k)^3$ terms on it imply the constraint $a_1=0$ for the ${\alpha'}^1$ term. Since there are no more cubic terms in ${\cal L}_{\mbox{eff}}$  there are no more constraints for the $a_j$ coefficients coming from the 3-point subamplitude.\\
Next, we consider the $4$-point subamplitude. Using the corresponding Feynman rules from ${\cal L}_{\mbox{eff}}$ in (\ref{L12simpler})  (with $a_1=0$)  it leads to
\begin{eqnarray}
A(1,2,3,4) & = & A_{YM}(1,2,3,4) +  \ g^2 \ (2 \alpha')^2 \biggl( \ a_3 {K^{(4)}_3}(1,2,3,4) + a_4 {K^{(4)}_4}(1,2,3,4) +\nonumber \\
&&
+ a_5 {K^{(4)}_5}(1,2,3,4) + a_6 {K^{(4)}_6}(1,2,3,4) \ \biggr) +   O({(2\alpha')}^{3}) \ ,
\label{A4}
\end{eqnarray}
\noindent where the ${K^{(4)}_j}(1,2,3,4)$'s ($j=3, 4, 5, 6$) are known $4$-point kinematical expressions given in Appendix \ref{4-point}.\\
In that appendix we see that demanding the absence of $(\zeta \cdot k)^4$ terms in the ${\alpha'}^2$ contribution to $A(1,2,3,4)$ implies that $a_3$, $a_4$, $a_5$ and $a_6$ should satisfy the following constraints:
\begin{eqnarray}
a_3 = -8 a_6 \ \ , a_4 = -4 a_6 \ \ , a_5 = 2 a_6 \ \ .
\label{coefficients1}
\end{eqnarray}
So, the conclusion of $step \ I$ of our procedure (see subsection \ref{How}) is that the only possible deformation of the bosonic part of the D=10 Super Yang-Mills lagrangian, allowed by Open Superstring Theory, is given by:
\begin{eqnarray}
{\cal L}_{\mbox{eff}} &=&  \frac{1}{g^2} \mbox{tr} \biggr[ -\frac{1}{4} F^{\mu\nu}F_{\mu\nu}
+ a_6 \ (2\alpha')^{2} \Big(-8 \ F^{\mu\lambda}F^{\nu}_{\ \lambda}F_{\mu}^{\ \rho}F_{\nu\rho}
 - \nonumber \\
&& \hphantom{   \ \mbox{tr} \biggr[ \ \  } 4  \ F^{\mu}_{\ \lambda}F_{\nu}^{\ \lambda}F^{\nu\rho}F_{\mu\rho}
+ 2 \ F^{\mu\nu}F_{\mu\nu}F^{\lambda\rho}F_{\lambda\rho} + \nonumber\\
&&
\hphantom{   \ \mbox{tr} \biggr[ \ \  }
F^{\mu\nu}F^{\lambda\rho}F_{\mu\nu}F_{\lambda\rho} \Big) + O({(2\alpha')}^{3}) \ \biggr] \ .
\label{L12simpler12}
\end{eqnarray}
\noindent Notice that this is in perfect agreement with the well known fact that D=10 SYM has a $unique$ deformation at ${\alpha'}^2$ order \cite{Cederwall2, Bergshoeff1}. Also, it is important to mention that the result we have arrived to in (\ref{L12simpler12}) is completely equivalent to the one
obtained by the method of BPS configurations \cite{Koerber1}.\\
\noindent Now we go to $step \ II$ of our procedure. Using the constraints in (\ref{coefficients1}) and the $4$-point kinematical expressions of  ${K^{(4)}_j}(1,2,3,4)$'s ($j=3, 4, 5, 6$) (see Appendix \ref{4-point}), comparison with the $4$-point subamplitude (\ref{A4fermionic}) at order ${\alpha'}^2$ (after using momentum conservation and the physical state condition)  leads to
\begin{eqnarray}
a_6 & = & -\frac{\pi^2}{96} \ ,
\label{a6}
\end{eqnarray}
\noindent which is the known ${\alpha'}^2$ correction to the D=10 Yang-Mills lagrangian coming from Open Superstring Theory \cite{Tseytlin2, Gross1}.\\

\subsection{Low energy effective lagrangian at ${\alpha'}^3$ order}

\label{Low3}

In Appendix \ref{alpha3} we show that the absence of $(\zeta \cdot k)^4$ terms in the $4$-point subamplitude of ${\cal L_{\mbox{eff}}}^{(3)}$, given in (\ref{L3}), implies that the coefficients of its $D^2F^4$ terms are constrained to satisfy
\begin{eqnarray}
- 2 a_{16} = -2 a_{17} = 8 a_{19} = - a_{20} = a_{22} \ ,  \nonumber \\
a_{18} = a_{21} = 0 \ .
\label{coefficients2}
\end{eqnarray}
Also in Appendix \ref{alpha3}, we show that the absence of $(\zeta \cdot k)^5$ terms in the $5$-point subamplitude of ${\cal L_{\mbox{eff}}}^{(3)}$ implies, besides (\ref{coefficients2}), that the remaining coefficients of its $F^5$ terms are constrained to satisfy
\begin{eqnarray}
a_{11} = a_{13} = -2 a_{15} =- i \ a_{22} \ ,\nonumber \\
a_{10} = a_{12} = a_{14} = 0 \ .
\label{coefficients3}
\end{eqnarray}
So, $step \ I$ of our procedure leads us to only one possible deformation of the bosonic part of the D=10 Super Yang-Mills lagrangian at order ${\alpha'}^3$, allowed by Open Superstring Theory:
\begin{eqnarray}
{\cal L_{\mbox{eff}}}^{(3)}&=& -  \frac{(2 \alpha')^3 \ a_{22}}{g^2} \ \mbox{tr} \biggl[ \  i \ F_{\mu}^{\ \nu}F_{\nu}^{\ \lambda}F_{\lambda}^{\ \rho}F_{\sigma}^{\ \mu}F_{\rho}^{\ \sigma}  + i \ F_{\mu}^{\ \nu}F_{\rho}^{\ \sigma}F_{\nu}^{\ \lambda}F_{\sigma}^{\ \mu}F_{\lambda}^{\ \rho} - \nonumber \\
& &\hphantom{   - a_{22} \frac{({2 \alpha'})^3}{g^2} \ \mbox{tr} \biggl[      }
\frac{i}{2} \ F_{\mu}^{\ \nu}F_{\nu}^{\ \lambda}F_{\rho}^{\ \sigma}F_{\lambda}^{\ \mu}F_{\sigma}^{\ \rho}   + \frac{1}{2} (D_{\mu}F_{\nu}^{\ \lambda})(D^{\mu}F_{\lambda}^{\ \rho})F_{\sigma}^{\ \nu}F_{\rho}^{\ \sigma}          \nonumber \\
& &\hphantom{    - a_{22} \frac{({2 \alpha'})^3}{g^2} \ \mbox{tr} \biggl[     }
\frac{1}{2} (D_{\mu}F_{\nu}^{\ \lambda})F_{\sigma}^{\ \nu}(D^{\mu}F_{\lambda}^{\ \rho})F_{\rho}^{\ \sigma}   -
\frac{1}{8} (D_{\mu}F_{\nu}^{\ \lambda})F_{\rho}^{\ \sigma}(D^{\mu}F_{\lambda}^{\ \nu})F_{\sigma}^{\ \rho}   + \nonumber \\
& &\hphantom{   - a_{22} \frac{({2 \alpha'})^3}{g^2} \ \mbox{tr} \biggl[      }  (D_{\sigma}F_{\mu}^{\ \nu})F_{\lambda}^{\ \rho}(D^{\mu}F_{\nu}^{\ \lambda})F_{\rho}^{\ \sigma} - F_{\mu}^{\ \nu}(D^{\mu}F_{\lambda}^{\ \rho})(D_{\sigma}F_{\nu}^{\ \lambda})F_{\rho}^{\ \sigma} \ \biggr] \ .
\label{L3final}
\end{eqnarray}
This result is also in perfect agreement with the fact that at ${\alpha'}^3$ there is a $unique$ supersymmetric deformation of the D=10 SYM lagrangian \cite{DeRoo2}. Koerber and Sevrin arrived to this same result (\ref{L3final}) in \cite{Koerber1}.\\
Demanding the 4-point subamplitude of the $D^2F^4$ terms in (\ref{L3final}) to agree with the corresponding open superstring 4-point amplitude (\ref{A4fermionic}) at ${\alpha'}^3$ order, leads us to
\begin{eqnarray}
a_{22}= 2 \ \zeta(3) \ .
\label{a22}
\end{eqnarray}
An interesting aspect, that we have verified in Appendix \ref{alpha3}, is that only demanding absence of $(\zeta \cdot k)^5$ terms in the $5$-point subamplitude of ${\cal L_{\mbox{eff}}}^{(3)}$ (and not worrying about the absence of the $(\zeta \cdot k)^4$ terms in the $4$-point subamplitude, as we did to obtain the relations in (\ref{coefficients3}))  is enough information to arrive to the whole set of relations in (\ref{coefficients2}) and (\ref{coefficients3}) and, therefore, to the expression of ${\cal L_{\mbox{eff}}}^{(3)}$ given in eq. (\ref{L3final}).\\
\begin{figure}[t]
\centerline{\includegraphics*[scale=0.4,angle=0]{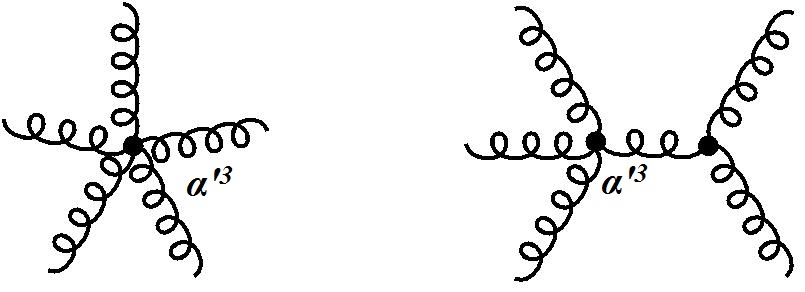}}
\caption{Feynman diagrams associated to the 5-point amplitude at ${\alpha'}^3$ order.}
\label{diagram5}
\end{figure}
\noindent The results in (\ref{L3final}) and (\ref{a22}) were first correctly obtained in \cite{Koerber1} for the D=10 case. They were confirmed by a $5$-point open superstring amplitude calculation in \cite{Brandt1}.

\section{The low energy effective lagrangian at ${\alpha'}^4$ order}

\label{Low}

${\alpha'}^4$ is the highest order for which $all$ the bosonic terms of the OSLEEL have been explicitly obtained in the literature up to this moment \cite{Koerber2}. Since the method that has been used to obtain these terms\footnote{The method of BPS configurations.} is not directly a String Theory one (like the conventional S matrix or the Sigma model methods) it is of much importance to see how the revisited S matrix method deals with them. We do this in the present section.\\
Since the calculations to obtain ${\cal L_{\mbox{eff}}}^{(4)}$ are quite extense, we will not present here the explicit list of a full basis of terms (with arbitrary coefficients) like we did in eqs. (\ref{L12simpler}) and (\ref{L3}), at lower orders in $\alpha'$. We will just mention that this basis contains 96 terms \cite{Koerber4}, we will present the final expression that we have obtained for ${\cal L_{\mbox{eff}}}^{(4)}$  and we will leave the details of the calculations to Appendix \ref{Some details}. \\
The expression that we have obtained for ${\cal L_{\mbox{eff}}}^{(4)}$ is the following:
\begin{eqnarray}
{\cal L_{\mbox{eff}}}^{(4)} & = & \frac{ (2 \alpha')^4 \pi^4}{g^2} \ \big( \ {\cal L}_{F^6} + {\cal L}_{D^2 F^5} + {\cal L}_{D^4 F^4} \ \big) \ ,
\label{L4}
\end{eqnarray}
where
\begin{eqnarray}
 \label{LF6}
\mathcal{L}_{F^{6}} &=&\frac{1}{46080} \ t_{(12)}^{\mu _{1}\nu _{1}\mu _{2}\nu _{2}\mu _{3}\nu
_{3}\mu _{4}\nu _{4}\mu _{5}\nu _{5}\mu _{6}\nu _{6}} \ \mbox{tr}\big(F_{\mu
_{1}\nu _{1}}F_{\mu _{2}\nu _{2}}F_{\mu _{3}\nu _{3}}F_{\mu _{4}\nu
_{4}}F_{\mu _{5}\nu _{5}}F_{\mu _{6}\nu _{6}}\big)\ , \\
\label{LD2F5}
\mathcal{L}_{D^{2}F^{5}} &=&\frac{56\ i}{46080}\ t_{(10)}^{\mu _{1}\nu
_{1}\mu _{2}\nu _{2}\mu _{3}\nu _{3}\mu _{4}\nu _{4}\mu _{5}\nu _{5}}\ %
\mbox{tr}\big(F_{\mu _{1}\nu _{1}}F_{\mu _{2}\nu _{2}}F_{\mu _{3}\nu
_{3}}D^{\alpha }F_{\mu _{4}\nu _{4}}D_{\alpha }F_{\mu _{5}\nu _{5}}\big)+
\nonumber \\
&&\frac{i}{46080}\ {(\eta \cdot t_{(8)})}^{\mu _{1}\nu _{1}\mu _{2}\nu
_{2}\mu _{3}\nu _{3}\mu _{4}\nu _{4}\mu _{5}\nu _{5}}\ \mbox{tr}\big( -169\
D^{\alpha }F_{\mu _{1}\nu _{1}}F_{\mu _{2}\nu _{2}}F_{\mu _{3}\nu
_{3}}F_{\mu _{4}\nu _{4}}D_{\alpha }F_{\mu _{5}\nu _{5}}+  \nonumber \\
&&\hphantom{ + \frac{i}{2880} \  {(\eta \cdot t_{(8)})_1}^{\mu _1 \nu _1 \mu
_2 \nu _2 \mu _3 \nu _3 \mu _4 \nu _4 \mu _5 \nu _5} \ \mbox{tr} \biggl[ }%
68\ D^{\alpha }F_{\mu _{1}\nu _{1}}D_{\alpha }F_{\mu _{2}\nu _{2}}F_{\mu
_{3}\nu _{3}}F_{\mu _{4}\nu _{4}}F_{\mu _{5}\nu _{5}}+  \nonumber \\
&&\hphantom{ + \frac{i}{2880} \  {(\eta \cdot t_{(8)})_1}^{\mu _1 \nu _1 \mu
_2 \nu _2 \mu _3 \nu _3 \mu _4 \nu _4 \mu _5 \nu _5} \ \mbox{tr} \biggl[ }%
237\ F_{\mu _{1}\nu _{1}}D^{\alpha }F_{\mu _{2}\nu _{2}}D_{\alpha }F_{\mu
_{3}\nu _{3}}F_{\mu _{4}\nu _{4}}F_{\mu _{5}\nu _{5}}+  \nonumber \\
&&\hphantom{ + \frac{i}{2880} \  {(\eta \cdot t_{(8)})_1}^{\mu _1 \nu _1 \mu
_2 \nu _2 \mu _3 \nu _3 \mu _4 \nu _4 \mu _5 \nu _5} \ \mbox{tr} \biggl[ }%
237\ F_{\mu _{1}\nu _{1}}D^{\alpha }F_{\mu _{2}\nu _{2}}F_{\mu _{3}\nu
_{3}}D_{\alpha }F_{\mu _{4}\nu _{4}}F_{\mu _{5}\nu _{5}}+  \nonumber \\
&&\hphantom{ + \frac{i}{2880} \  {(\eta \cdot t_{(8)})_1}^{\mu _1 \nu _1 \mu
_2 \nu _2 \mu _3 \nu _3 \mu _4 \nu _4 \mu _5 \nu _5} \ \mbox{tr} \biggl[ }%
267\ F_{\mu _{1}\nu _{1}}F_{\mu _{2}\nu _{2}}D^{\alpha }F_{\mu _{3}\nu
_{3}}D_{\alpha }F_{\mu _{4}\nu _{4}}F_{\mu _{5}\nu _{5}}+  \nonumber \\
&&\hphantom{ + \frac{i}{2880} \  {(\eta \cdot t_{(8)})_1}^{\mu _1 \nu _1 \mu
_2 \nu _2 \mu _3 \nu _3 \mu _4 \nu _4 \mu _5 \nu _5} \ \mbox{tr} \biggl[ }%
16\ F_{\mu _{1}\nu _{1}}F_{\mu _{2}\nu _{2}}F_{\mu _{3}\nu _{3}}D^{\alpha
}F_{\mu _{4}\nu _{4}}D_{\alpha }F_{\mu _{5}\nu _{5}}\ \big)-  \nonumber \\
&& \frac{i}{5760}~  t_{(8)}^{\mu _{1}\nu _{1}\mu _{2}\nu _{2}\mu
_{3}\nu _{3}\mu _{4}\nu _{4}}\  \Big\{ 17 \ \mbox{tr}\big(D^{\mu _{5}}F_{\mu _{1}\nu
_{1}}F_{\mu _{2}\nu _{2}}F_{\mu _{3}\nu _{3}} D^{\nu
_{5}}F_{\mu _{4}\nu _{4}}F_{\mu _{5}\nu _{5}}\big) + \nonumber \\
&&\hphantom{ -\ \frac{i}{5760}~  t_{(8)}^{\mu _{1}\nu _{1}\mu _{2}\nu _{2}\mu
_{3}\nu _{3}\mu _{4}\nu _{4}}\     } +2 \ \mbox{tr}\big( F_{\mu _{1}\nu _{1}} D^{\mu _{5}}F_{\mu _{2}\nu _{2}}D^{\nu _{5}}F_{\mu
_{3}\nu _{3}}F_{\mu _{4}\nu _{4}}  F_{\mu _{5}\nu
_{5}} \big)  \nonumber \Big\}\ ,  \\
\label{LD4F4}
\mathcal{L}_{D^{4}F^{4}} &=&-\frac{1}{11520}\ t_{(8)}^{\mu _{1}\nu _{1}\mu
_{2}\nu _{2}\mu _{3}\nu _{3}\mu _{4}\nu _{4}}\ \mbox{tr}\big( D^{\alpha
}F_{\mu _{1}\nu _{1}}D_{(\alpha }D_{\beta )}F_{\mu _{2}\nu _{2}}D^{\beta
}F_{\mu _{3}\nu _{3}}F_{\mu _{4}\nu _{4}}+  \nonumber \\
&&\hphantom{ - \frac{1}{1440} t_{(8)}^{\mu_1 \nu_1 \mu_2 \nu_2 \mu_3 \nu_3
\mu_4 \nu_4} \ \mbox{tr} \big[ }+8\ D^{\alpha }F_{\mu _{1}\nu _{1}}D_{\alpha
}F_{\mu _{2}\nu _{2}}D^{\beta }F_{\mu _{3}\nu _{3}}D_{\beta }F_{\mu _{4}\nu
_{4}}\ \big) \ .
\end{eqnarray}
\noindent As expected,  a new 12-index tensor $t_{(12)}$ (characteristic of $6$-point scattering) has arisen. Its explicit expression, as a sum of products of 6 $\eta_{\mu \nu}$'s, can be obtained from formula (\ref{t12}) in appendix \ref{t12tensor}. $\eta \cdot t_{(8)}$ and $t_{(10)}$  are 10-index tensors that already appeared in our expression for the open superstring 5-point amplitude \cite{Barreiro1}. In Appendices \ref{t8tensor} and \ref{t10tensor} we recall how to construct them, respectively. $t_{(8)}$ is, of course, the well known $8$-index tensor that appears in $4$ open superstring scattering \cite{Schwarz1}. In Appendix \ref{t8tensor} we also recall how to construct it. All these tensors are antisymmetric under the interchange of indices $\mu_i$ and $\nu_i$.\\
It is quite remarkable that we have obtained all coefficients and terms in the lagrangian in (\ref{L4}) without using $any$ scattering amplitude information from Open Superstring Theory at ${\alpha'}^4$ order (not even the $4$-point amplitude). We have just demanded the $(\zeta \cdot k)^N$ terms to be absent in the $N$-point amplitude of the general lagrangian at ${\alpha'}^4$ order (with $N=4,5,6$) and this has fixed all its coefficients. One might think that the best scenario could have been that this last condition fixed the lagrangian coefficients up to a global factor, as it happended with the ${\alpha'}^2$ and ${\alpha'}^3$ order contributions (see eqs. (\ref{L12simpler12}) and (\ref{L3final}), respectively) and then it should have been necessary to use information from the open superstring $4$-point amplitude at ${\alpha'}^4$ order. But what in fact happened is that even the global coefficient has now been fixed by the condition of absence of $(\zeta \cdot k)^6$ terms in the $6$-point amplitude\footnote{Since the coefficients of $all$ ${\alpha'}^4$ terms have been determined, that is the reason of why in the second column of the table in eq. (\ref{table2}), for $p=4$ we have written that the dimension of the constrained basis is $0$.}. The reason for this is that, at ${\alpha'}^4$ order, the $6$-point amplitude not only receives contributions from Feynman diagrams constructed with the Yang-Mills propagator, the Yang-Mills vertices and the ${\alpha'}^4$ order vertices (and these last ones contain the originally unknown $a_j$ coefficients), but it receives as well contributions from diagrams which contain the ${\alpha'}^2$ order vertices (which coefficients are all known and proportional to $\pi^2$). So the linear system of equations for the unknown coefficients of ${\cal L_{\mbox{eff}}}^{(4)}$ is not homogeneous and it happens to have a $unique$ solution, which leads to our result in eq. (\ref{L4}) (see more details in Appendix \ref{Some details}).\\
\begin{figure}[t]
\centerline{\includegraphics*[scale=0.4,angle=0]{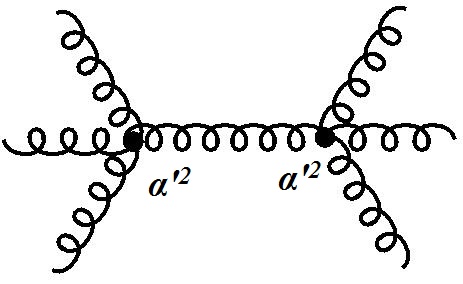}}
\caption{Feynman diagram that contributes to the 6-point amplitude at ${\alpha'}^4$ order.
This diagram is responsible for fixing all the $a_j$ coefficients of
${\cal  L}_{\mbox{eff}}^{(4)}$ without using any explicit information from Open Superstring Theory.}
\label{diagramc}
\end{figure}
\noindent The fact of finding all the ${\alpha'}^4$ terms of the low energy effective lagrangian without needing to use any information from Open Superstring Theory at that $\alpha'$ order also happened in the method of BPS configurations \cite{Koerber2, Koerber4}. It means that ${\cal L_{\mbox{eff}}}^{(4)}$ (and its fermionic completion) is the ${\alpha'}^4$ supersymmetric deformation of ${\cal L}_{SYM} + (2 \alpha')^2 {\cal L}^{(2)}$ (where ${\cal L}^{(2)}$ is the known first supersymmetric correction to the D=10 SYM lagrangian \cite{Cederwall1, Cederwall2}). In fact, in \cite{Howe1} it was proved that this ${\alpha'}^4$ supersymmetric correction should exist (but it was not computed explicitly) and ${\cal L_{\mbox{eff}}}^{(4)}$ should match the bosonic part of the one given by the algorithm of that reference.   \\
In Appendix \ref{Some details} we have verified that our lagrangian in (\ref{L4}) bypasses the following tests:
\begin{enumerate}
\item The abelian limit of the ${\cal L}_{F^6}$ agrees with the corresponding $F^6$ terms of the (supersymmetric) Born-Infeld lagrangian \cite{Tseytlin2, DeRoo3}.
\item The $5$-point amplitude obtained from ${\cal L}_{D^2 F^5} + {\cal L}_{D^4 F^4}$ agrees $exactly$ with the corresponding one coming from Open Superstring Theory, obtained by us in \cite{Barreiro1}, at ${\alpha'}^4$ order.
\item  The $4$-point amplitude obtained from ${\cal L}_{D^4 F^4}$ agrees with the corresponding one obtained from the open superstring one, eq.(\ref{A4fermionic}),  at ${\alpha'}^4$ order. In particular, this means that the abelian limit of ${\cal L}_{D^4 F^4}$ should agree with the $\partial^4 F^4$ terms of \cite{DeRoo3, Andreev1} since those terms agree with the abelian $4$-point amplitude of Open Superstring Theory.
\end{enumerate}
These tests guarantee that our expression for ${\cal L_{\mbox{eff}}}^{(4)}$ is correct up to terms which are sensible to 6 and higher $n$-point amplitudes. They also confirm that the abelian 6-point amplitude that comes from our low energy effective lagrangian is correct at ${\alpha'}^4$ order. But
there is still an additional nonabelian test for the $F^6$ terms, which is extremely important and comes from the very nature of our revisited S-matrix method. This test makes direct contact with Open Superstring Theory calculations at ${\alpha'}^4$ order: the 6-point amplitude at ${\alpha'}^4$ order that can be computed from  ${\cal L_{\mbox{eff}}}$, expanded up to ${\alpha'}^4$ order, where ${\cal L_{\mbox{eff}}}^{(4)}$ is the one that we have found in eq. (\ref{L4}) (with the corresponding expressions in (\ref{LF6}), (\ref{LD2F5}) and (\ref{LD4F4})), and the corresponding amplitude from Open Superstring Theory, agree in the fact that both of them have no $(\zeta \cdot k)^6$ terms. At this point, we claim that our expression for ${\cal L_{\mbox{eff}}}^{(4)}$ is in $complete$ agreement with the nonabelian $6$-point amplitude at ${\alpha'}^4$ order because of the $uniqueness$ of the expression that we have found for it (up to terms which are not sensible to S-matrix calculations)\footnote{And this expression has been obtained by means of an S-matrix method.}. It $should$ happen that the remaining $(\zeta \cdot k)^4(\zeta \cdot \zeta)$ and $(\zeta \cdot k)^2 (\zeta \cdot \zeta)^2$ terms  of the nonabelian 6-point amplitude agree with the ones that come from the corresponding open superstring 6-point amplitude or otherwise the $N$-point formula in eq.(\ref{ANfermionic}) would be incorrect in the case of $N=6$ (and we clearly do not believe this since the vertex operator formalism in Open Superstring Theory leads to expressions for the scattering amplitudes which respect basic properties like unitarity and gauge invariance). \\
It would be nice to verify the equivalence between the ${\alpha'}^4$ terms found by the method of BPS configurations \cite{Koerber2}, ours and the ones proposed by the algorithm in \cite{Howe1}, but this would require huge additional calculations to be done. Koerber and Sevrin present their ${\alpha'}^4$  order result in terms of symmetrized traces and commutators of covariant derivatives, which is in fact a very compact way of writing the terms, but it is different from our way of presenting the result in eq. (\ref{L4}). Our main worry has not been to write the final answer in a short manner, but to write the terms in such a way that it is clear which ones are sensible at least to 4, 5 and 6-point amplitudes. We just mention here that the result of Koerber and Sevrin \cite{Koerber2} also satisfies the test mentioned in item $1$ and the abelian part of the test in item $3$ and that their method is self consistent in the sense of finding a $unique$ solution for a linear system of equations (for the coefficients in ${\cal L_{\mbox{eff}}}^{(4)}$, for a given basis of terms) which is overdetermined.\\
\noindent  In Appendix \ref{Some details} we comment on the possibility of arriving to our result in eqs. (\ref{L4}) - (\ref{LD4F4}), much in the same spirit that me mentioned at the end of subsection \ref{Low3}, that is, by just demanding the absense of $(\zeta \cdot k)^6$ terms in the $6$-point amplitude, while not imposing the absense of $(\zeta \cdot k)^N$ terms in the $4$ and $5$-point amplitudes.

\section{Final remarks and future prospects}

\label{Summary}

In this work we have presented a `revisited' S-matrix approach to obtain the bosonic terms of the open superstring low energy effective lagrangian (OSLEEL). It is just the well known S-matrix approach, but we have called it `revisited' because we have found a specific way of using the information contained in open superstring scattering amplitudes, in such a way that the calculations are simpler than the corresponding ones in the bosonic theory of open strings. In order to obtain the OSLEEL ${\alpha'}^p$ order terms, the method usually proceeds in two steps\footnote{Here we have said `usually' because at ${\alpha'}^4$ order the first step is enough to determine all the coefficients of the OSLEEL, but for any other $\alpha'$ order we expect both steps to be required.}. The first step consists in demanding the absence of $(\zeta \cdot k)^N$ terms
in the $N$-point (tree level) gauge boson subamplitude (constraint in eq. (\ref{constraint})), in the field theory side, for $N=4, \ldots, p+2$\footnote{Here we have two observations:
\begin{enumerate}
\item The method applies also when $N=3$, but the only case in which this has any relevance at all is when proving that there are no ${\alpha'}^1$ terms in the OSLEEL.
\item We have found that for the $5$-point subamplitude (at ${\alpha'}^3$ order), demanding this constraint just for $N=p+2$ is enough, but we are not sure if this also happens for the 6-point subamplitude at  ${\alpha'}^4$ order.
\end{enumerate}}. Using these constraints at each ${\alpha'}^p$ order, this step reduces enormously the number of unkowns coefficients in the OSLEEL (in comparison to the number of coefficients existing in the general lagrangian; for example of this comparison see the table in (\ref{table2})). The remaining unknowns are determined in the second step of the method by matching the field theory $N$-point amplitude with information from the $N$-point amplitude of Open Superstring Theory at ${\alpha'}^p$ order (where $N$ is expected to be much lower than $p+2$; in fact, at least up to ${\alpha'}^4$ order terms, we have been able to obtain the OSLEEL only using the $N=4$ scattering amplitude of open superstrings). This is the main difference with the `conventional' S-matrix approach, in which it is generally believed that an open superstring $(p+2)$-point amplitude is needed (at least expanded at ${\alpha'}^p$ order) in order to fully determine the ${\alpha'}^{p}$ order terms of the low energy effective lagrangian (where $p \ge 2$).\\
Although we have not further studied the ${\alpha'}^p$ order terms for $p \ge 5$, we expect our revisited S-matrix approach to be capable, in principle, to determine completely the OSLEEL. We will examine more carefully this issue on a forthcoming work \cite{Barreiro3} and we will also examine there the possibility of the kinematical constraint in eq. (\ref{constraint}) being valid in $any$ theory which consistently considers supersymmetric deformations of D=10 SYM theory, not only Open Superstring Theory.\\
The sort of restrictions that arise for the coefficients of the bosonic terms of the OSLEEL, in this `revisited' S-matrix method, are similar to the ones found in the method of BPS configurations (due to Koerber and Sevrin \cite{Koerber1, Koerber2, Koerber3}\footnote{This method was firstly applied to the abelian case in \cite{Fosse1}.}). Both methods agree in their results up to ${\alpha'}^3$ order and probably agree at ${\alpha'}^4$ order, as well (see section \ref{Low} for more details about this last comparison).\\
The constraint (\ref{constraint}), which is the crucial part of our `revisited' S-matrix approach, seems not to exist in Open Bosonic String Theory\footnote{At least in the case of $3$ and $4$-point subamplitudes it is known not to happen and it is clear that in the $N$-point subamplitude formula \cite{Schwarz1}  the $(\zeta \cdot k)^N$ terms will show up, but we have not checked that for $N \ge 5$ those terms do not cancel each other.}. This suggests that the reason for it is Spacetime Supersymmetry (an inherent symmetry of Open Superstring Theory) and that is why the `conventional' S-matrix approach (which treats the bosonic and the supersymmetric theory of open strings, at least in the determination of the bosonic terms of the OSLEEL, on the same footing) does not include it. \\
We end this work by mentioning that there is a natural extension of the requirement (\ref{constraint}) and its consequences for the OSLEEL, to the case of the Closed Superstring Low Energy Effective Lagrangean (CSLEEL) of the NS-NS sector of the Type II Theories. This can easily be understood by means of the KLT relations \cite{Kawai1}. A careful analysis of the kinematics involved in these relations tells us that demanding the $(\zeta \cdot k)^N$ terms to be absent in gauge boson $N$-point amplitudes of Open Superstring Theory implies that in the interactions of gravitons and Kalb-Ramond states, in the Type II theories\footnote{We have made some change in eq.(\ref{constraint2}), with respect to the first version of this work that we sent to the hep-th Arxive, due to an observation by M. R. Garousi. In the way that we have now written this equation, it is consistent for $N=4$ with the forbidden kinematical terms mentioned by him, after eq. (23) of \cite{Garousi1}.},
\begin{eqnarray}
\mbox{the} N-\mbox{point amplitude contains neither}  (k \ \zeta \ k)^{N}  nor (k \ \zeta \ \zeta \ k)^1 \ (k \ \zeta \ k)^{N-2} \mbox{ terms} \ . \nonumber \\
\label{constraint2}
\end{eqnarray}
For example, if we consider the $3$-point amplitude of gravitons and/or Kalb-Ramond states, the constraints in (\ref{constraint2}) imply that terms like
\begin{eqnarray}
(k_2 \ \zeta_1 \ k_2)(k_3 \ \zeta_2 \ \zeta_3 \ k_1) & = & ({k_2}_{\mu} {\zeta_1}^{\mu \nu} {k_2}_{\nu}) ({k_3}_{\rho} {\zeta_2}^{\rho \sigma}  {\zeta_3}_{\sigma \lambda} {k_1}^{\lambda})  \ , \\
(k_2 \ \zeta_1 \ k_2)(k_3 \ \zeta_2 \ k_3)(k_1 \ \zeta_3 \ k_1) & = & ({k_2}_{\mu} {\zeta_1}^{\mu \nu} {k_2}_{\nu}) ({k_3}_{\rho} {\zeta_2}^{\rho \sigma} {k_3}_{\sigma}) ({k_1}_{\lambda} {\zeta_3}^{\lambda \omega} {k_1}_{\omega}) \ ,
\end{eqnarray}
should not appear in the amplitude. It can be easily checked that these terms do indeed appear in the $3$-point amplitude in the case of Closed Bosonic String Theory, at ${\alpha'}^1$ and ${\alpha'}^2$ order, respectively. The fact that they should be absent in the supersymmetric case implies that there should be no ${\alpha'}^1$ and no ${\alpha'}^2$ terms in the NS-NS sector of the low energy effective lagrangian of the Type II String Theories, as it is well known.\\
The remaining dependence of the NS-NS sector of the CSLEEL can be simply obtained recalling the result of ref. \cite{Gross2}, where it was shown that, starting from the low energy effective lagrangian of the pure gravitational sector of the Type II theories, the dependence of it on the dilaton ($\phi$) and the Kalb-Ramond field ($B_{\mu \nu}$) can be directly inferred from the first one by just replacing the curvature tensor by the combination
\begin{eqnarray}
{\bar R}_{\mu\nu}{}^{\lambda\rho}=R_{\mu\nu}{}^{\lambda\rho}+
 e^{-\Phi} \ \nabla_{[\mu}
H_{\nu]}{}^{\lambda\rho}-\delta_{[\mu}{}^{[\lambda}\nabla_{\nu]}
\nabla^{\rho]}\Phi \ , \label{Rbar}
\end{eqnarray}
where $H_{\mu \nu \lambda} =\partial_{[\mu} B_{\nu \lambda]}$.\\
So, if our revisited S-matrix approach is indeed capable of determining the complete OSLEEL, it is likely to happen the same thing with the NS-NS sector of Closed Superstring Theory.

\section*{Acknowledgements}

We would like to thank M. de Roo for carefully reading and making comments to this work and also for very useful e-mail correspondence.  R. M. would like to thank A. Tseytlin and M. R. Garousi for useful e-mail correspondence and especially  P. Howe and P. Koerber for many detailed explanations.

\appendix

\section{Conventions and identities}
\label{conventions}

\subsection{Metric, symmetrization and antisymmetrization over spacetime indexes:}

We use the following convention for the Minkowski metric:
\begin{eqnarray}
\label{metric} \eta_{\mu \nu} = \mbox{diag}(-, +, \ldots , +) \ .
\end{eqnarray}
The symmetrization and antisymmetrization convention that we use,
on the spacetime indexes of a product of two  vectors $A$ and $B$,
is the following:
\begin{eqnarray}
\label{symmetrization} A^{(\mu}B^{\nu)}=\frac{1}{2} (
A^{\mu}B^{\nu}+A^{\nu}B^{\mu}) \ , \\
\label{antisymmetrization} A^{[\mu}B^{\nu]}=\frac{1}{2} (
A^{\mu}B^{\nu}-A^{\nu}B^{\mu} ) \ .
\end{eqnarray}

\subsection{Gauge group generators, field strength and covariant derivative:}

Gauge fields are matrices in the Lie group internal space, so that
$A_{\mu} = A^{\mu}_{\ a} \lambda^a$, where the $\lambda^a$ are the
generators in the adjoint representation,
\begin{eqnarray}
(\lambda^a)^{bc} = -i f^{abc} \ ,
\label{adjointm}
\end{eqnarray}
which satisfy the normalization relation
\begin{eqnarray}
\mbox{tr}(\lambda^a \lambda^b) = \delta^{ab} \ .\label{group1}
\end{eqnarray}
The field strength and the covariant derivative are defined by\footnote{In contrast to the conventions that we used in \cite{Barreiro1}, now the coupling constant $g$ does not come in the definitions (\ref{group2}) and (\ref{group3}), neither in the identity (\ref{group4}): it comes as a global $1/g^2$ factor in the whole low energy effective lagrangian. See, for example, eqs.(\ref{L12simpler}) and (\ref{L3}).}
\begin{eqnarray}
\label{group2} F_{\mu \nu} & = & \partial_{\mu} A_{\nu} -
\partial_{\nu} A_{\mu}
- i [ A_{\mu}, A_{\nu}] \ , \\
\label{group3} D_{\mu} \phi & = & \partial_{\mu} \phi - i [
A_{\mu}, \phi ] \ ,
\end{eqnarray}
and they are related by the identity
\begin{eqnarray}
[D_{\mu}, D_{\nu}] \phi & = & -i \ [ F_{\mu \nu}, \phi ] \ .
\label{group4}
\end{eqnarray}
Covariant derivatives of field strengths satisfy the Bianchi
identity:
\begin{eqnarray}
\label{Bianchi} D^{\mu} F^{\nu \rho} + D^{\rho} F^{\mu \nu} +
D^{\nu} F^{\rho \mu} = 0 \ .
\end{eqnarray}

\section{Vertices and tensors}
\label{tensors}

\subsection{Yang-Mills vertices}
\vspace{-0.45cm}
\begin{eqnarray}
\label{YMvertex3}
V^{(3)}_{YM \ \mu_1 \mu_2 \mu_3}(k_1, k_2, k_3)
&=& -i \ [\frac{}{} \eta_{\mu_1 \mu_2} (k_1-k_2)_{\mu_3} +
\eta_{\mu_2 \mu_3}
(k_2-k_3)_{\mu_1} + \eta_{\mu_3 \mu_1} (k_3-k_1)_{\mu_2} ] \ , \nonumber \\
\label{YMvertex4} \\
V^{(4)}_{YM \ \mu_1 \mu_2 \mu_3 \mu_4}(k_1, k_2,
k_3, k_4 ) &=& - \ [\frac{}{} \eta_{\mu_1 \mu_2} \eta_{\mu_3
\mu_4} - 2 \eta_{\mu_1 \mu_3} \eta_{\mu_2 \mu_4} + \eta_{\mu_4
\mu_1} \eta_{\mu_2 \mu_3} \frac{}{}] \ .
\end{eqnarray}

\subsection{$t_{(8)}$ and $\eta \cdot t_{(8)}$ tensors}
\label{t8tensor}

\noindent The $t_{(8)}$ tensor\footnote{An explicit expression for
it may be found in equation (4.A.21) of \cite{Schwarz1}.},
characteristic of the 4 boson scattering amplitude, is
antisymmetric on each pair $(\mu_j ,\nu_j)$ ($j=1,2,3,4$) and is
symmetric under any exchange of such of pairs.  It satisfies the
identity\footnote{Formula (\ref{t8}) has been taken from appendix
A of \cite{DeRoo1}.}:
\begin{multline}
t^{(8)}_{\mu_1 \nu_1 \mu_2 \nu_2 \mu_3 \nu_3 \mu_4 \nu_4}
A_1^{\mu_1 \nu_1} A_2^{\mu_2 \nu_2} A_3^{\mu_3 \nu_3} A_4^{\mu_4
\nu_4}  = \\
\begin{split}
&  -2 \biggl( \mbox{Tr}(A_1 A_2)\mbox{Tr}(A_3 A_4) + \mbox{Tr}(A_1
A_3)\mbox{Tr}(A_2 A_4) +
\mbox{Tr}(A_1 A_4)\mbox{Tr}(A_2 A_3) \biggr)+ \\
& + 8 \biggl( \mbox{Tr}(A_1 A_2 A_3 A_4) + \mbox{Tr}(A_1 A_3 A_2
A_4) + \mbox{Tr}(A_1 A_3 A_4 A_2)  \biggr) \ ,
\end{split}
\label{t8}
\end{multline}
where the $A_j$ tensors are antisymmetric and where `Tr' means
the trace over the spacetime indexes.\\
\noindent A ten index tensor, which is also antisymmetric on each
pair $(\mu_j ,\nu_j)$, can be constructed from the Minkowski
metric tensor and the $t_{(8)}$ one, as follows:
\begin{eqnarray}
{(\eta \cdot t_{(8)})}^{\mu_1 \nu_1 \mu_2 \nu_2 \mu_3 \nu_3
\mu_4 \nu_4 \mu_5 \nu_5} & = & \eta^{\nu_3 \nu_4} t_{(8)}^{\mu_3
\mu_4 \mu_5 \nu_5 \mu_1 \nu_1 \mu_2 \nu_2} + \eta^{\mu_3 \mu_4}
t_{(8)}^{\nu_3 \nu_4 \mu_5 \nu_5 \mu_1 \nu_1 \mu_2 \nu_2} -
\nonumber \\
&&- \ \eta^{\mu_3 \nu_4} t_{(8)}^{\nu_3 \mu_4 \mu_5 \nu_5 \mu_1
\nu_1 \mu_2 \nu_2} - \eta^{\nu_3 \mu_4} t_{(8)}^{\mu_3 \nu_4 \mu_5
\nu_5 \mu_1 \nu_1 \mu_2 \nu_2} \ . \label{etat8}
\end{eqnarray}
This tensor appears in the 5-point amplitude of the open
superstring \cite{Barreiro1}\footnote{In ref. \cite{Barreiro1} we used a subindex `1' for the $\eta \cdot t_{(8)}$ tensor, as a reminder that the twisting relation that it obeys, eq. (\ref{etat8twisting}), is realized with respect to the vertex `1' on the disk, but on this paper we have omitted that reminder and written that tensor with no subindex.}. It also changes sign under a twisting
transformation\footnote{See the third item of subsection 4.2 of
\cite{Barreiro1} for further details about a twisting transformation
on the disk.} with respect to index 1, that is,
\begin{eqnarray}
{(\eta \cdot t_{(8)})}^{\mu_1 \nu_1 \mu_5 \nu_5 \mu_4 \nu_4
\mu_3 \nu_3 \mu_2 \nu_2} = - {(\eta \cdot t_{(8)})}^{\mu_1 \nu_1
\mu_2 \nu_2 \mu_3 \nu_3 \mu_4 \nu_4 \mu_5 \nu_5} \ .
\label{etat8twisting}
\end{eqnarray}
It satisfies an identity similar to the one given in (\ref{t8}) for $t_{(8)}$:
\begin{eqnarray}
&&(\eta \cdot  t_{(8)})_{\mu _{1}\nu _{1}\mu _{2}\nu _{2}\mu _{3}\nu _{3}\mu _{4}\nu
_{4}\mu _{5}\nu _{5}}A_{1}^{\mu _{1}\nu _{1}}A_{2}^{\mu _{2}\nu
_{2}}A_{3}^{\mu _{3}\nu _{3}}A_{4}^{\mu _{4}\nu _{4}}A_{5}^{\mu _{5}\nu
_{5}}=  \nonumber \\
&&\hphantom{+++}+2\left[ \ \ \mbox{Tr}\left( A_{1}A_{2}\right) \mbox{Tr}\left(
A_{3}A_{4}A_{5}\right) +\mbox{Tr}\left( A_{2}A_{5}\right) \mbox{Tr}\left(
A_{1}A_{3}A_{4}\right) +\mbox{Tr}\left( A_{1}A_{5}\right) \mbox{Tr}\left(
A_{2}A_{3}A_{4}\right) \ \ \right] -  \nonumber \\
&&\hphantom{+++}-6\left[ \ \ \mbox{Tr}\left( A_{1}A_{5}\right) \mbox{Tr}\left(
A_{2}A_{4}A_{3}\right) +\mbox{Tr}\left( A_{1}A_{2}\right) \mbox{Tr}\left(
A_{3}A_{5}A_{4}\right) +\mbox{Tr}\left( A_{2}A_{5}\right) \mbox{Tr}\left(
A_{1}A_{4}A_{3}\right) \ \ \right] +  \nonumber \\
&&\hphantom{+++}+4\left[ \ \ \mbox{Tr}\left( A_{1}A_{2}A_{4}A_{3}A_{5}\right) -%
\mbox{Tr}\left( A_{1}A_{2}A_{3}A_{4}A_{5}\right) -\mbox{Tr}\left(
A_{1}A_{2}A_{5}A_{3}A_{4}\right) \right. +  \nonumber \\
&&\hphantom{++++.}+\left. \mbox{Tr}\left( A_{1}A_{2}A_{5}A_{4}A_{3}\right) -%
\mbox{Tr}\left( A_{1}A_{3}A_{4}A_{2}A_{5}\right) +\mbox{Tr}\left(
A_{1}A_{4}A_{3}A_{2}A_{5}\right) \ \ \right] +  \nonumber \\
&&\hphantom{++.}+12\left[ \ \ \mbox{Tr}\left( A_{1}A_{4}A_{3}A_{5}A_{2}\right) -%
\mbox{Tr}\left( A_{1}A_{3}A_{4}A_{5}A_{2}\right) -\mbox{Tr}\left(
A_{1}A_{5}A_{2}A_{3}A_{4}\right) \right. +  \nonumber \\
&&\hphantom{++++.}+\left. \mbox{Tr}\left( A_{1}A_{5}A_{2}A_{4}A_{3}\right) -%
\mbox{Tr}\left( A_{1}A_{5}A_{3}A_{4}A_{2}\right) +\mbox{Tr}\left(
A_{1}A_{5}A_{4}A_{3}A_{2}\right) \ \ \right]   \ .
\label{nt8}
\end{eqnarray}

\subsection{$t_{(10)}$ tensor}
\label{t10tensor}

The $t_{(10)}$ tensor is another ten index tensor that appears in
the 5-point amplitude of the open superstring \cite{Barreiro1}. It is linearly
independent to the $(\eta \cdot t_{(8)})_1$ one. This tensor also satisfies an identity
similar to the one given in (\ref{t8}) for $t_{(8)}$:
\vspace{-0.2cm}
\begin{multline}
t^{(10)}_{\mu_1 \nu_1 \mu_2 \nu_2 \mu_3 \nu_3 \mu_4 \nu_4 \mu_5
\nu_5} A_1^{\mu_1 \nu_1} A_2^{\mu_2 \nu_2} A_3^{\mu_3 \nu_3}
A_4^{\mu_4 \nu_4} A_5^{\mu_5 \nu_5} = \\
\begin{split}
& \hphantom{+ } - 8 \left[  \ \ \mbox{Tr}(A_1 A_2)\mbox{Tr}(A_3
A_4 A_5) + \mbox{Tr}(A_1 A_3)\mbox{Tr}(A_2 A_4 A_5) +
\mbox{Tr}(A_1 A_4)\mbox{Tr}(A_2 A_3 A_5) + \right. \\
& \hphantom{+ \ \ \ 8} + \mbox{Tr}(A_1 A_5)\mbox{Tr}(A_2 A_3 A_4)
+ \mbox{Tr}(A_2 A_3)\mbox{Tr}(A_1 A_4 A_5) +
\mbox{Tr}(A_2 A_4)\mbox{Tr}(A_1 A_3 A_5) + \\
& \hphantom{+ \ \ \ 8} + \mbox{Tr}(A_2 A_5)\mbox{Tr}(A_1 A_3 A_4)
+ \mbox{Tr}(A_3 A_4)\mbox{Tr}(A_1 A_2 A_5) +
\mbox{Tr}(A_3 A_5)\mbox{Tr}(A_1 A_2 A_4) + \\
& \hphantom{+ \ \ \ 8} \left. + \mbox{Tr}(A_4 A_5)\mbox{Tr}(A_1
A_2 A_3) \ \ \right] + \ 48 \ \mbox{Tr}(A_1 A_2 A_3 A_4 A_5) \ +
\\
&+ 16 \left[ \ \ \mbox{Tr}(A_1 A_2 A_3 A_5 A_4) + \mbox{Tr}(A_1
A_2 A_4 A_3 A_5) + \mbox{Tr}(A_1 A_2 A_5 A_3 A_4) + \right. \\
& \hphantom{+ \ 16}+ \mbox{Tr}(A_1 A_2 A_4 A_5 A_3) -
\mbox{Tr}(A_1 A_2 A_5 A_4 A_3) + \mbox{Tr}(A_1 A_3 A_2 A_4 A_5) -
\\
& \hphantom{+ \ 16}- \mbox{Tr}(A_1 A_3 A_2 A_5 A_4) +
\mbox{Tr}(A_1 A_4 A_2 A_3 A_5) + \mbox{Tr}(A_1 A_5 A_2 A_3 A_4) -
\\
& \hphantom{+ \ 16} \left. - \mbox{Tr}(A_1 A_4 A_2 A_5 A_3) -
\mbox{Tr}(A_1 A_5 A_2 A_4 A_3) \ \ \right] \ ,
\end{split}
\label{t10}
\end{multline}
where the $A_j$ fields are antisymmetric. From (\ref{t10}) an
explicit expression of the $t_{(10)}$ tensor may be obtained, once
its symmetry properties are considered.

\subsection{$t_{(12)}$ tensor}

\label{t12tensor}

\noindent Consider the 12-index tensor $s_{(12)}$ which is obtained by means of the relation
\vspace{-0.2cm}
\begin{eqnarray}
&&s_{\mu _{1}\nu _{1}\mu _{2}\nu _{2}\mu _{3}\nu _{3}\mu _{4}\nu _{4}\mu
_{5}\nu _{5}\mu _{6}\nu _{6}}^{(12)} A_{1}^{\mu _{1}\nu _{1}}A_{2}^{\mu
_{2}\nu _{2}}A_{3}^{\mu _{3}\nu _{3}}A_{4}^{\mu _{4}\nu _{4}}A_{5}^{\mu
_{5}\nu _{5}}A_{6}^{\mu _{6}\nu _{6}}= \hspace{5cm} \nonumber \\
&&\hphantom{+} -144 \ \mbox{Tr}\left( A_{1} A_{6}\right) \mbox{Tr}\left( A_{2} A_{5}\right) \mbox{Tr}\left(
A_{3} A_{4}\right) -1396 \ \mbox{Tr}\left( A_{1} A_{2} A_{5} A_{6}\right) \mbox{Tr}\left(
A_{3} A_{4}\right) +  \nonumber \\
&& \hphantom{+}+2260 \ \mbox{Tr}\left( A_{1} A_{5} A_{2} A_{6}\right) \mbox{Tr}\left( A_{3} A_{4}\right)
-2016 \ \mbox{Tr}\left( A_{1} A_{5} A_{6} A_{2}\right) \mbox{Tr}\left( A_{3} A_{4}%
\right) +  \nonumber \\
&& \hphantom{+}+4028 \ \mbox{Tr}\left( A_{1} A_{6} A_{2} A_{5}\right) \mbox{Tr}\left( A_{3} A_{4}\right)
-2172 \ \mbox{Tr}\left( A_{1} A_{6} A_{5} A_{2}\right) \mbox{Tr}\left( A_{3} A_{4}%
\right) + \nonumber \\
&& \hphantom{+}+104 \ \mbox{Tr}\left( A_{1} A_{4}\right) \mbox{Tr}\left( A_{2} A_{6}\right) \mbox{Tr}\left(
A_{3} A_{5}\right) -80 \ \mbox{Tr}\left( A_{1} A_{4}\right) \mbox{Tr}\left( A_{2} A_{5}\right)
\mbox{Tr}\left( A_{3} A_{6}\right) -  \nonumber \\
&& \hphantom{+}-64 \ \mbox{Tr}\left( A_{1} A_{6}\right) \mbox{Tr}\left( A_{2} A_{3}\right) \mbox{Tr}\left(
A_{4} A_{5}\right) +304 \ \mbox{Tr}\left( A_{1} A_{5}\right) \mbox{Tr}\left( A_{2} A_{3}%
\right) \mbox{Tr}\left( A_{4} A_{6}\right) +  \nonumber \\
&& \hphantom{+}+300 \ \mbox{Tr}\left( A_{1} A_{3} A_{5}\right) \mbox{Tr}\left( A_{2} A_{4} A_{6}\right)
 -180 \ \mbox{Tr}\left( A_{1} A_{3} A_{5}\right) \mbox{Tr}\left( A_{2} A_{6} A_{4}\right)-  \nonumber \\
&& \hphantom{+}-1210 \ \mbox{Tr}\left( A_{1} A_{2} A_{6}\right) \mbox{Tr}\left( A_{3} A_{4} A_{5}\right)
 +696 \ \mbox{Tr}\left( A_{1} A_{2} A_{6}\right) \mbox{Tr}\left( A_{3} A_{5} A_{4}\right)
 +  \nonumber \\
&& \hphantom{+}+220 \ \mbox{Tr}\left( A_{1} A_{2} A_{4}\right) \mbox{Tr}\left( A_{3} A_{5} A_{6}\right)
-4692 \ \mbox{Tr}\left( A_{1} A_{4} A_{2}\right) \mbox{Tr}\left( A_{3} A_{5} A_{6}%
\right) -  \nonumber \\
&& \hphantom{+}-660 \ \mbox{Tr}\left( A_{1} A_{2} A_{4}\right) \mbox{Tr}\left( A_{3} A_{6} A_{5}\right)
+1980 \ \mbox{Tr}\left( A_{1} A_{4} A_{2}\right) \mbox{Tr}\left( A_{3} A_{6} A_{5}%
\right) -  \nonumber \\
&& \hphantom{+}-4032 \ \mbox{Tr}\left( A_{1} A_{3} A_{2}\right) \mbox{Tr}\left( A_{4} A_{5} A_{6}\right)
-4316 \ \mbox{Tr}\left( A_{4} A_{6}\right) \mbox{Tr}\left( A_{1} A_{2} A_{3} A_{5}\right) +  \nonumber \\
&& \hphantom{+}+534 \ \mbox{Tr}\left( A_{3} A_{6}\right) \mbox{Tr}\left( A_{1} A_{2} A_{5} A_{4}\right)
 +1602 \ \mbox{Tr}\left( A_{3} A_{6}\right) \mbox{Tr}\left( A_{1} A_{4} A_{5} A_{2}%
\right) - \nonumber \\
&& \hphantom{+}-294 \ \mbox{Tr}\left( A_{2} A_{5}\right) \mbox{Tr}\left( A_{1} A_{3} A_{4} A_{6}\right)
-3124 \ \mbox{Tr}\left( A_{2} A_{4}\right) \mbox{Tr}\left( A_{1} A_{3} A_{5} A_{6}%
\right) +  \nonumber \\
&& \hphantom{+}-1228 \ \mbox{Tr}\left( A_{2} A_{3}\right) \mbox{Tr}\left( A_{1} A_{4} A_{5} A_{6}\right)
 + 3684 \ \mbox{Tr}\left( A_{2} A_{3}\right) \mbox{Tr}\left( A_{1} A_{6} A_{4} A_{5}\right)+  \nonumber \\
&& \hphantom{+}+1228 \ \mbox{Tr}\left( A_{2} A_{4}\right) \mbox{Tr}\left( A_{1} A_{5} A_{3} A_{6}\right)
+1228 \ \mbox{Tr}\left( A_{2} A_{3}\right) \mbox{Tr}\left( A_{1} A_{5} A_{4} A_{6}%
\right) +  \nonumber \\
&& \hphantom{+}+3684 \ \mbox{Tr}\left( A_{2} A_{4}\right) \mbox{Tr}\left( A_{1} A_{6} A_{3} A_{5}\right)
-882 \ \mbox{Tr}\left( A_{2} A_{5}\right) \mbox{Tr}\left( A_{1} A_{6} A_{4} A_{3}%
\right) -  \nonumber \\
&& \hphantom{+}-3684 \ \mbox{Tr}\left( A_{2} A_{3}\right) \mbox{Tr}\left( A_{1} A_{6} A_{5} A_{4}\right)
-144 \ \mbox{Tr}\left( A_{1} A_{6}\right) \mbox{Tr}\left( A_{2} A_{3} A_{5} A_{4}%
\right) -  \nonumber \\
&& \hphantom{+}-432 \ \mbox{Tr}\left( A_{1} A_{6}\right) \mbox{Tr}\left( A_{2} A_{4} A_{5} A_{3}\right)
+8240 \ \mbox{Tr}\left( A_{1} A_{2} A_{3} A_{4} A_{5} A_{6}\right) +  \nonumber \\
&& \hphantom{+}+7680 \ \mbox{Tr}\left( A_{1} A_{2} A_{4} A_{3} A_{5} A_{6}\right) -8256 \ \mbox{Tr}%
\left( A_{1} A_{2} A_{5} A_{6} A_{3} A_{4}\right) +2624 \ \mbox{Tr}\left(
A_{1} A_{2} A_{6} A_{4} A_{3} A_{5}\right) \nonumber +
\end{eqnarray}

\begin{eqnarray}
&& \hphantom{+}+9824 \ \mbox{Tr}\left( A_{1} A_{3} A_{2} A_{4} A_{5} A_{6}\right) -432 \ \mbox{Tr}\left(
A_{1} A_{3} A_{2} A_{4} A_{6} A_{5}\right) -7840 \ \mbox{Tr}\left(
A_{1} A_{3} A_{2} A_{5} A_{6} A_{4}\right) + \nonumber \\
&& \hphantom{+} +9824 \ \mbox{Tr}\left( A_{1} A_{3} A_{4} A_{2} A_{5} A_{6}\right)
+4032 \ \mbox{Tr}\left(
A_{1} A_{3} A_{4} A_{5} A_{6} A_{2}\right)
-4032 \ \mbox{Tr}\left(
A_{1} A_{3} A_{4} A_{6} A_{5} A_{2}\right) - \nonumber \\
&& \hphantom{+}
-256 \ \mbox{Tr}\left( A_{1} A_{3} A_{5} A_{2} A_{4} A_{6}\right)
+1120 \ \mbox{Tr}\left(
A_{1} A_{3} A_{5} A_{2} A_{6} A_{4}\right) +4032 \ \mbox{Tr}\left(
A_{1} A_{3} A _{5} A_{6} A _{4} A_{2}\right) - \nonumber \\
&& \hphantom{+}
-384 \ \mbox{Tr}\left( A_{1} A_{3} A _{6} A_{2} A_{5} A_{4}\right) -4032 \ \mbox{Tr}\left(
A_{1} A_{3} A _{6} A_{5} A_{4} A_{2}\right)
+9824 \ \mbox{Tr}\left(
A_{1} A_{4} A_{2} A_{3} A_{5} A_{6}\right) + \nonumber \\
&& \hphantom{+}
+9824 \ \mbox{Tr}\left( A_{1} A _{4} A_{3} A_{2} A_{5} A_{6}\right) +4032 \ \mbox{Tr}\left(
A_{1} A_{4} A_{3} A_{5} A_{6} A_{2}\right)+4032 \ \mbox{Tr}\left(
A_{1} A_{4} A_{5} A_{6} A_{3} A_{2}\right) - \nonumber \\
&& \hphantom{+} -4032 \ \mbox{Tr}\left(
A_{1} A_{4} A_{6} A_{5} A_{3} A_{2}\right)
-928 \ \mbox{Tr}\left(
A_{1} A_{5} A_{2} A_{3} A_{4} A_{6}\right) -9824 \ \mbox{Tr}\left(A_{1} A_{5} A_{2} A_{4} A_{3} A_{6}\right) -  \nonumber \\
&& \hphantom{+} -9824 \ \mbox{Tr}\left(
A_{1} A_{5} A_{3} A_{2} A_{4} A_{6}\right) -9824 \ \mbox{Tr}\left(
A_{1} A_{5} A_{3} A_{4} A_{2} A_{6}\right) -9824 \ \mbox{Tr}\left(
A_{1} A_{5} A_{4} A_{2} A_{3} A_{6}\right) - \nonumber \\
&& \hphantom{+}  -9824 \ \mbox{Tr}\left(
A_{1} A_{5} A_{4} A_{3} A_{2} A_{6}\right) +4032 \ \mbox{Tr}\left(A_{1} A_{5} A_{6} A_{3} A_{4} A_{2}\right) +4032 \ \mbox{Tr}\left(
A_{1} A_{5} A_{6} A_{4} A_{3} A_{2}\right) , \nonumber \\
\label{t12}
\end{eqnarray}
where the $A_j$ tensors are antisymmetric ($j=1, \ldots, 6$). \\
The $t_{(12)}$ tensor appearing in eq. (\ref{LF6}) is given by an averaged expression of the $s_{(12)}$, such that the resulting tensor obeys the twisting relation (\ref{t12F6}) of Appendix \ref{Some details}:
\begin{eqnarray}
t_{\mu _{1}\nu _{1}\mu _{2}\nu _{2}\mu _{3}\nu _{3}\mu _{4}\nu _{4}\mu
_{5}\nu _{5}\mu _{6}\nu _{6}}^{(12)}\hspace{-0.2pt} & = & \hspace{-0.2pt} \frac{1}{2}\  \Big( \ s_{\mu _{1}\nu _{1}\mu _{2}\nu _{2}\mu _{3}\nu _{3}\mu _{4}\nu _{4}\mu
_{5}\nu _{5}\mu _{6}\nu _{6}}^{(12)}\hspace{-0.2pt}+\hspace{-0.2pt}s_{\mu _{1}\nu _{1}\mu _{6}\nu _{6}\mu _{5}\nu _{5}\mu _{4}\nu _{4}\mu
_{3}\nu _{3}\mu _{2}\nu _{2}}^{(12)}  \ \Big) \ .
\end{eqnarray}

\section{The $4$-point kinematical factor}
\label{Gauge boson}

The $4$-point kinematic factor appearing in eq.(\ref{A4fermionic}) is given by \cite{Schwarz1}
\begin{eqnarray}
K &= & - \frac{1}{4} \Bigl[ ts(\zeta_1 \cdot \zeta_3)(\zeta_2 \cdot
  \zeta_4) + su(\zeta_2 \cdot \zeta_3)(\zeta_1 \cdot \zeta_4) +
  ut(\zeta_1 \cdot \zeta_2)(\zeta_3 \cdot \zeta_4) \Bigr] +
\nonumber\\
&&{}+ \frac{1}{2} s \Bigl[ (\zeta_1 \cdot k_4)(\zeta_3 \cdot
  k_2)(\zeta_2 \cdot \zeta_4) + (\zeta_2 \cdot k_3)(\zeta_4 \cdot
  k_1)(\zeta_1 \cdot \zeta_3) +
\nonumber \\&&
\hphantom{{}+ \frac{1}{2} s \Bigl[}
+(\zeta_1 \cdot k_3)(\zeta_4 \cdot
  k_2)(\zeta_2 \cdot \zeta_3)
+ (\zeta_2 \cdot k_4)(\zeta_3 \cdot k_1)(\zeta_1 \cdot \zeta_4)
  \Bigr] +
\nonumber\\ &&
{}+ \frac{1}{2} t \Bigl[ (\zeta_2 \cdot k_1)(\zeta_4 \cdot
  k_3)(\zeta_3 \cdot \zeta_1) + (\zeta_3 \cdot k_4)(\zeta_1 \cdot
  k_2)(\zeta_2 \cdot \zeta_4) +
\nonumber \\&&
\hphantom{{}+ \frac{1}{2} t \Bigl[}
+ (\zeta_2 \cdot k_4)(\zeta_1 \cdot k_3)(\zeta_3 \cdot \zeta_4) +
  (\zeta_3 \cdot k_1)(\zeta_4 \cdot k_2)(\zeta_2 \cdot \zeta_1)
  \Bigr] +
\nonumber\\ &&
{}+\frac{1}{2} u \Bigl[ (\zeta_1 \cdot k_2)(\zeta_4 \cdot
  k_3)(\zeta_3 \cdot \zeta_2) +
 (\zeta_3
  \cdot k_4)(\zeta_2 \cdot k_1)(\zeta_1 \cdot \zeta_4) +
\nonumber \\&&
\hphantom{{}+\frac{1}{2} u \Bigl[}
+  (\zeta_1 \cdot k_4)(\zeta_2 \cdot k_3)(\zeta_3 \cdot \zeta_4) +
  (\zeta_3 \cdot k_2)(\zeta_4 \cdot k_1)(\zeta_1 \cdot \zeta_2)
  \Bigr]
\label{Ks}
\end{eqnarray}
and where $\{ s, t, u \}$ are the Mandelstam variables defined in (\ref{Mandelstam}). \\
Notice that the expression for $K$ in (\ref{Ks}) contains no $(\zeta \cdot k)^4$ terms, but does not have manifest (on-shell) gauge invariance.
An alternative expression for it, which now has this symmetry in a manifest way (due to the symmetries of the $t_{(8)}$ tensor), is the following \cite{Schwarz1}:
\begin{eqnarray}
K &= & t_{(8)}^{\mu_1 \nu_1 \mu_2 \nu_2 \mu_3 \nu_3 \mu_4 \nu_4}
\zeta^1_{\mu_1} k^1_{\nu_1} \zeta^2_{\mu_2} k^2_{\nu_2}
\zeta^3_{\mu_3} k^3_{\nu_3} \zeta^4_{\mu_4} k^4_{\nu_4} \ .
\label{manifest1}
\end{eqnarray}
When explicitly expanded it becomes
\begin{eqnarray}
K &= & \frac{1}{2} \ \Big[
 -(\zeta_1 \cdot \zeta_4) (\zeta_3 \cdot k_1) (\zeta_2 \cdot k_4) (k_2 \cdot k_3)+(\zeta_4 \cdot k_1) (\zeta_1 \cdot \zeta_3)
 (\zeta_2 \cdot k_4) (k_2 \cdot k_3) -\nonumber \\
&& \hphantom{  \frac{1}{2} \ \Big[  }   -(k_1 \cdot k_4) (\zeta_1 \cdot \zeta_3) (\zeta_2 \cdot \zeta_4) (k_2 \cdot k_3) +(\zeta_4 \cdot k_1) (\zeta_1 \cdot k_3) (k_2 \cdot k_4) (\zeta_2 \cdot \zeta_3)+\nonumber
\end{eqnarray}
\begin{eqnarray}
&& \hphantom{  \frac{1}{2} \ \Big[  } +(\zeta_1 \cdot \zeta_4) (k_1 \cdot k_3) (\zeta_2 \cdot k_4)
 (\zeta_3 \cdot k_2)+(\zeta_1 \cdot \zeta_4) (\zeta_3 \cdot k_1) (k_2 \cdot k_4) (\zeta_2 \cdot k_3)-\nonumber \\
&& \hphantom{  \frac{1}{2} \ \Big[  } -(\zeta_1 \cdot k_4) ) (\zeta_3 \cdot k_1) (\zeta_4 \cdot k_2) (\zeta_2 \cdot k_3)+(k_1 \cdot k_4) (\zeta_1 \cdot \zeta_3) (\zeta_4 \cdot k_2) (\zeta_2 \cdot k_3)+\nonumber \\
&& \hphantom{  \frac{1}{2} \ \Big[  } +(\zeta_1 \cdot k_4) (\zeta_3 \cdot k_1) (\zeta_2 \cdot \zeta_4) (k_2 \cdot k_3)+(k_1 \cdot k_4) (\zeta_1 \cdot k_3) (\zeta_2 \cdot \zeta_4) (\zeta_3 \cdot k_2)+\nonumber \\
&& \hphantom{  \frac{1}{2} \ \Big[  } +(\zeta_1 \cdot k_4) (k_1 \cdot k_3) (\zeta_4 \cdot k_2) (\zeta_2 \cdot \zeta_3)-(k_1 \cdot k_4) (\zeta_1 \cdot k_3) (\zeta_4 \cdot k_2) (\zeta_2 \cdot \zeta_3)-\nonumber \\
&& \hphantom{  \frac{1}{2} \ \Big[  } -(\zeta_4 \cdot k_1) (\zeta_1 \cdot k_3) (\zeta_2 \cdot k_4) (\zeta_3 \cdot k_2)-(\zeta_1 \cdot \zeta_4) (k_1 \cdot k_3) (k_2 \cdot k_4) (\zeta_2 \cdot \zeta_3)-\nonumber \\
&& \hphantom{  \frac{1}{2} \ \Big[  } -(\zeta_3 \cdot k_1) (\zeta_2 \cdot \zeta_4) (\zeta_1 \cdot k_3) (k_2 \cdot k_4)+(\zeta_1 \cdot \zeta_3) (\zeta_2 \cdot \zeta_4) (k_1 \cdot k_3) (k_2 \cdot k_4)-\nonumber \\
&& \hphantom{  \frac{1}{2} \ \Big[  } -(\zeta_1 \cdot \zeta_3) (\zeta_4 \cdot k_2) (k_1 \cdot k_3) (\zeta_2 \cdot k_4)+(\zeta_3 \cdot k_1) (\zeta_4 \cdot k_2)
  (\zeta_1 \cdot k_3) (\zeta_2 \cdot k_4)-\nonumber \\
&& \hphantom{  \frac{1}{2} \ \Big[  } -(\zeta_4 \cdot k_1) (\zeta_1 \cdot k_4) (k_2 \cdot k_3) (\zeta_2 \cdot \zeta_3)-(\zeta_1 \cdot \zeta_4) (k_1 \cdot k_4) (\zeta_2 \cdot k_3) (\zeta_3 \cdot k_2)-\nonumber \\
&& \hphantom{  \frac{1}{2} \ \Big[  } -(\zeta_4 \cdot k_1) (\zeta_1 \cdot \zeta_3) (k_2 \cdot k_4) (\zeta_2 \cdot k_3)-(\zeta_1 \cdot k_4) (k_1 \cdot k_3) (\zeta_2 \cdot \zeta_4) (\zeta_3 \cdot k_2)-\nonumber \\
&& \hphantom{  \frac{1}{2} \ \Big[  } -(k_1 \cdot k_2) (\zeta_2 \cdot \zeta_4) (\zeta_1 \cdot \zeta_3) (k_3 \cdot k_4)-(k_1 \cdot k_4) (k_2 \cdot k_3) (\zeta_1 \cdot \zeta_2) (\zeta_3 \cdot \zeta_4)-\nonumber \\
&& \hphantom{  \frac{1}{2} \ \Big[  } -(\zeta_4 \cdot k_1) (\zeta_2 \cdot k_3) (\zeta_1 \cdot k_2) (\zeta_3 \cdot k_4)+(\zeta_4 \cdot k_1) (\zeta_2 \cdot \zeta_3) (\zeta_1 \cdot k_2) (k_3 \cdot k_4)-\nonumber \\
&& \hphantom{  \frac{1}{2} \ \Big[  } -(\zeta_1 \cdot \zeta_4) (\zeta_2 \cdot \zeta_3) (k_1 \cdot k_2) (k_3 \cdot k_4)-
(\zeta_4 \cdot k_1) (\zeta_3 \cdot k_2) (\zeta_1 \cdot \zeta_2) (k_3 \cdot k_4)+\nonumber \\
&& \hphantom{  \frac{1}{2} \ \Big[  } +(k_1 \cdot k_4) (\zeta_3 \cdot k_2) (\zeta_1 \cdot \zeta_2) (\zeta_4 \cdot k_3)+(\zeta_1 \cdot \zeta_4) (\zeta_2 \cdot k_3) (k_1 \cdot k_2) (\zeta_3 \cdot k_4)+\nonumber \\
&& \hphantom{  \frac{1}{2} \ \Big[  } +(\zeta_4 \cdot k_1) (k_2 \cdot k_3) (\zeta_1 \cdot \zeta_2) (\zeta_3 \cdot k_4)-(\zeta_1 \cdot \zeta_4) (k_2 \cdot k_3) (\zeta_2 \cdot k_1) (\zeta_3 \cdot k_4)+\nonumber \\
&& \hphantom{  \frac{1}{2} \ \Big[  } +(\zeta_1 \cdot k_4) (k_2 \cdot k_3) (\zeta_2 \cdot k_1) (\zeta_3 \cdot \zeta_4)-(k_1 \cdot k_4) (\zeta_2 \cdot \zeta_3) (\zeta_1 \cdot k_2) (\zeta_4 \cdot k_3)+\nonumber \\
&& \hphantom{  \frac{1}{2} \ \Big[  } +(\zeta_1 \cdot k_4) (\zeta_2 \cdot \zeta_3) (k_1 \cdot k_2) (\zeta_4 \cdot k_3)
 -(\zeta_1 \cdot k_4) (\zeta_3 \cdot k_2) (\zeta_2 \cdot k_1) (\zeta_4 \cdot k_3)+\nonumber \\
&& \hphantom{  \frac{1}{2} \ \Big[  } +(k_1 \cdot k_4) (\zeta_2 \cdot k_3) (\zeta_1 \cdot k_2) (\zeta_3 \cdot \zeta_4)-(\zeta_1 \cdot k_4) (\zeta_2 \cdot k_3) (k_1 \cdot k_2) (\zeta_3 \cdot \zeta_4)+\nonumber \\
&& \hphantom{  \frac{1}{2} \ \Big[  } +(\zeta_1 \cdot \zeta_4) (\zeta_3 \cdot k_2) (\zeta_2 \cdot k_1) (k_3 \cdot k_4)+(\zeta_4 \cdot k_1) (\zeta_1 \cdot k_4) (\zeta_2 \cdot k_3)
 (\zeta_3 \cdot k_2)+\nonumber \\
&& \hphantom{  \frac{1}{2} \ \Big[  } +(\zeta_1 \cdot \zeta_4) (k_1 \cdot k_4) (k_2 \cdot k_3) (\zeta_2 \cdot \zeta_3)-(k_1 \cdot k_2) (\zeta_1 \cdot \zeta_2) (\zeta_3 \cdot k_4) (\zeta_4 \cdot k_3)-\nonumber \\
&& \hphantom{  \frac{1}{2} \ \Big[  } -(\zeta_1 \cdot k_2) (\zeta_2 \cdot k_1) (k_3 \cdot k_4) (\zeta_3 \cdot \zeta_4)+(k_1 \cdot k_2) (\zeta_1 \cdot \zeta_2) (k_3 \cdot k_4) (\zeta_3 \cdot \zeta_4)+\nonumber \\
&& \hphantom{  \frac{1}{2} \ \Big[  } +(\zeta_1 \cdot k_2) (\zeta_2 \cdot k_1) (\zeta_3 \cdot k_4) (\zeta_4 \cdot k_3)+(\zeta_2 \cdot k_1) (\zeta_4 \cdot k_2) (\zeta_1 \cdot \zeta_3) (k_3 \cdot k_4)-\nonumber \\
&& \hphantom{  \frac{1}{2} \ \Big[  } -(\zeta_1 \cdot k_2) (\zeta_2 \cdot k_4) (\zeta_3 \cdot k_1) (\zeta_4 \cdot k_3)-(k_1 \cdot k_2) (\zeta_2 \cdot k_4) (\zeta_1 \cdot k_3) (\zeta_3 \cdot \zeta_4)+\nonumber \\
&& \hphantom{  \frac{1}{2} \ \Big[  } +(\zeta_1 \cdot k_2) (\zeta_2 \cdot k_4) (k_1 \cdot k_3) (\zeta_3 \cdot \zeta_4)+(\zeta_2 \cdot k_1) (k_2 \cdot k_4) (\zeta_1 \cdot k_3) (\zeta_3 \cdot \zeta_4)-\nonumber \\
&& \hphantom{  \frac{1}{2} \ \Big[  } -(\zeta_1 \cdot \zeta_2) (k_2 \cdot k_4) (k_1 \cdot k_3) (\zeta_3 \cdot \zeta_4)-(\zeta_2 \cdot k_1) (\zeta_4 \cdot k_2) (\zeta_1 \cdot k_3) (\zeta_3 \cdot k_4)+\nonumber \\
&& \hphantom{  \frac{1}{2} \ \Big[  } +(\zeta_1 \cdot \zeta_2) (\zeta_4 \cdot k_2) (k_1 \cdot k_3) (\zeta_3 \cdot k_4)+(k_1 \cdot k_2) (\zeta_2 \cdot k_4) (\zeta_1 \cdot \zeta_3) (\zeta_4 \cdot k_3)-\nonumber \\
&& \hphantom{  \frac{1}{2} \ \Big[  } -(\zeta_2 \cdot k_1) (k_2 \cdot k_4) (\zeta_1 \cdot \zeta_3) (\zeta_4 \cdot k_3)+(k_1 \cdot k_2) (\zeta_2 \cdot \zeta_4) (\zeta_1 \cdot k_3) (\zeta_3 \cdot k_4)-\nonumber \\
&& \hphantom{  \frac{1}{2} \ \Big[  } -(\zeta_1 \cdot k_2) (\zeta_2 \cdot \zeta_4) (k_1 \cdot k_3) (\zeta_3 \cdot k_4)+(\zeta_1 \cdot k_2)
 (\zeta_2 \cdot \zeta_4) (\zeta_3 \cdot k_1) (k_3 \cdot k_4)-\nonumber \\
&& \hphantom{  \frac{1}{2} \ \Big[  } -(\zeta_1 \cdot \zeta_2)  (\zeta_4 \cdot k_2)(\zeta_3 \cdot k_1) (k_3 \cdot k_4)+ (\zeta_1 \cdot \zeta_2) (k_2 \cdot k_4) (\zeta_3 \cdot k_1)(\zeta_4 \cdot k_3) \ \Big] \ .
\label{manifest2}
\end{eqnarray}
Notice that this expression does contain $(\zeta \cdot k)^4$ terms\footnote{See, for example, the second term in the ninth line of eq.(\ref{manifest2}).}, but the one in (\ref{Ks}) does not. This is just a mathematical artifact, in order to implement (on-shell) gauge symmetry in a manifest way: when using momentum conservation and the physical state conditions all $(\zeta \cdot k)^4$ terms disappear in (\ref{manifest2}).

\section{${\alpha'}^2$ kinematic calculations}
\label{4-point}

Here we present the 4-point kinematical expressions introduced on equation (\ref{A4})
\vspace{-2 cm}
\begin{eqnarray}
K_{3}^{(4)} &=&2(\zeta _{1}.\zeta _{3})(\zeta _{2}.\zeta _{4})\alpha _{12}^{2}+(\zeta _{1}.\zeta _{4})(\zeta _{2}.k_{1})(\zeta _{3}.k_{1})\alpha _{12}+(\zeta _{1}.\zeta _{4})(\zeta _{2}.k_{3})(\zeta _{3}.k_{1})\alpha _{12}-  \nonumber \\
&&-2(\zeta _{1}.k_{2})(\zeta _{2}.\zeta _{4})(\zeta _{3}.k_{1})\alpha _{12}+2(\zeta _{1}.k_{3})(\zeta _{2}.\zeta _{4})(\zeta _{3}.k_{2})\alpha _{12}-(\zeta _{1}.k_{3})(\zeta _{2}.k_{1})(\zeta _{3}.\zeta _{4})\alpha _{12}-  \nonumber \\
&&-(\zeta _{1}.k_{3})(\zeta _{2}.k_{3})(\zeta _{3}.\zeta _{4})\alpha _{12}-2(\zeta _{1}.\zeta _{3})(\zeta _{2}.k_{1})(\zeta _{4}.k_{1})\alpha _{12}-2(\zeta _{1}.\zeta _{3})(\zeta _{2}.k_{3})(\zeta _{4}.k_{1})\alpha _{12}-  \nonumber \\
&&-2(\zeta _{1}.\zeta _{3})(\zeta _{2}.k_{1})(\zeta _{4}.k_{2})\alpha _{12}-(\zeta _{1}.k_{3})(\zeta _{2}.\zeta _{3})(\zeta _{4}.k_{2})\alpha _{12}+(\zeta _{1}.\zeta _{2})(\zeta _{3}.k_{1})(\zeta _{4}.k_{2})\alpha _{12}+  \nonumber \\
&&+2(\zeta _{1}.\zeta _{3})(\zeta _{2}.\zeta _{4})\alpha _{13}\alpha _{12}+(\zeta _{1}.\zeta _{4})(\zeta _{2}.\zeta _{3})\alpha _{13}^{2}+(\zeta _{1}.\zeta _{3})(\zeta _{2}.\zeta _{4})\alpha _{13}^{2}+(\zeta _{1}.\zeta _{2})(\zeta _{3}.\zeta _{4})\alpha _{13}^{2}+
\nonumber \\
&&+(\zeta _{1}.k_{2})(\zeta _{2}.k_{1})(\zeta _{3}.k_{1})(\zeta
_{4}.k_{1})+(\zeta _{1}.k_{2})(\zeta _{2}.k_{3})(\zeta _{3}.k_{1})(\zeta
_{4}.k_{1})-(\zeta _{1}.k_{3})(\zeta _{2}.k_{1})(\zeta _{3}.k_{2})(\zeta
_{4}.k_{1})-  \nonumber \\
&&-(\zeta _{1}.k_{3})(\zeta _{2}.k_{3})(\zeta _{3}.k_{2})(\zeta
_{4}.k_{1})+(\zeta _{1}.k_{2})(\zeta _{2}.k_{1})(\zeta _{3}.k_{1})(\zeta
_{4}.k_{2})-(\zeta _{1}.k_{3})(\zeta _{2}.k_{1})(\zeta _{3}.k_{1})(\zeta
_{4}.k_{2})-  \nonumber \\
&&-(\zeta _{1}.k_{3})(\zeta _{2}.k_{3})(\zeta _{3}.k_{1})(\zeta
_{4}.k_{2})-(\zeta _{1}.k_{3})(\zeta _{2}.k_{1})(\zeta _{3}.k_{2})(\zeta
_{4}.k_{2})+(\zeta _{1}.\zeta _{4})(\zeta _{2}.k_{1})(\zeta
_{3}.k_{1})\alpha _{13}-  \nonumber \\
&&-2(\zeta _{1}.k_{2})(\zeta _{2}.\zeta _{4})(\zeta _{3}.k_{1})\alpha _{13}-(\zeta _{1}.k_{3})(\zeta _{2}.\zeta _{4})(\zeta _{3}.k_{1})\alpha _{13}+(\zeta _{1}.\zeta _{4})(\zeta _{2}.k_{1})(\zeta _{3}.k_{2})\alpha _{13}+  \nonumber \\
&&+(\zeta _{1}.\zeta _{4})(\zeta _{2}.k_{3})(\zeta _{3}.k_{2})\alpha _{13}-2(\zeta _{1}.k_{2})(\zeta _{2}.\zeta _{4})(\zeta _{3}.k_{2})\alpha _{13}+(\zeta _{1}.k_{2})(\zeta _{2}.k_{1})(\zeta _{3}.\zeta _{4})\alpha _{13}-  \nonumber \\
&&-(\zeta _{1}.k_{3})(\zeta _{2}.k_{1})(\zeta _{3}.\zeta _{4})\alpha _{13}+(\zeta _{1}.k_{2})(\zeta _{2}.k_{3})(\zeta _{3}.\zeta _{4})\alpha _{13}-2(\zeta _{1}.\zeta _{3})(\zeta _{2}.k_{1})(\zeta _{4}.k_{1})\alpha _{13}-  \nonumber \\
&&-(\zeta _{1}.k_{3})(\zeta _{2}.\zeta _{3})(\zeta _{4}.k_{1})\alpha _{13}+(\zeta _{1}.\zeta _{2})(\zeta _{3}.k_{1})(\zeta _{4}.k_{1})\alpha _{13}-(\zeta _{1}.\zeta _{3})(\zeta _{2}.k_{1})(\zeta _{4}.k_{2})\alpha _{13}+  \nonumber \\
&&+(\zeta _{1}.\zeta _{3})(\zeta _{2}.k_{3})(\zeta _{4}.k_{2})\alpha _{13}+(\zeta _{1}.k_{2})(\zeta _{2}.\zeta _{3})(\zeta _{4}.k_{2})\alpha _{13}+2(\zeta _{1}.\zeta _{2})(\zeta _{3}.k_{1})(\zeta _{4}.k_{2})\alpha _{13}+  \nonumber \\
&&+(\zeta _{1}.\zeta _{2})(\zeta _{3}.k_{2})(\zeta _{4}.k_{2})\alpha _{13}
\end{eqnarray}%
\vspace{-1.4 cm}
\begin{eqnarray}
K_{4}^{(4)}= &&(\zeta _{1}.\zeta _{4})(\zeta _{2}.\zeta _{3})\alpha _{12}^{2}+(\zeta _{1}.\zeta _{2})(\zeta _{3}.\zeta _{4})\alpha _{12}^{2}+(\zeta _{1}.\zeta _{4})(\zeta _{2}.k_{1})(\zeta _{3}.k_{1})\alpha _{12}+(\zeta _{1}.\zeta _{4})(\zeta _{2}.k_{3})(\zeta _{3}.k_{1})\alpha _{12}+  \nonumber \\
&&+(\zeta _{1}.\zeta _{4})(\zeta _{2}.k_{3})(\zeta _{3}.k_{2})\alpha _{12}-(\zeta _{1}.k_{2})(\zeta _{2}.k_{1})(\zeta _{3}.\zeta _{4})\alpha _{12}-(\zeta _{1}.k_{3})(\zeta _{2}.k_{1})(\zeta _{3}.\zeta _{4})\alpha _{12}-  \nonumber \\
&&-(\zeta _{1}.k_{3})(\zeta _{2}.k_{3})(\zeta _{3}.\zeta _{4})\alpha _{12}-(\zeta _{1}.k_{2})(\zeta _{2}.\zeta _{3})(\zeta _{4}.k_{1})\alpha _{12}-(\zeta _{1}.k_{3})(\zeta _{2}.\zeta _{3})(\zeta _{4}.k_{1})\alpha _{12}-  \nonumber \\
&&-(\zeta _{1}.\zeta _{2})(\zeta _{3}.k_{1})(\zeta _{4}.k_{1})\alpha _{12}-(\zeta _{1}.\zeta _{2})(\zeta _{3}.k_{2})(\zeta _{4}.k_{1})\alpha _{12}-(\zeta _{1}.k_{3})(\zeta _{2}.\zeta _{3})(\zeta _{4}.k_{2})\alpha _{12}-  \nonumber \\
&&-(\zeta _{1}.\zeta _{2})(\zeta _{3}.k_{2})(\zeta _{4}.k_{2})\alpha _{12}+2(\zeta _{1}.\zeta _{2})(\zeta _{3}.\zeta _{4})\alpha _{13}\alpha _{12}+(\zeta _{1}.\zeta _{2})(\zeta _{3}.\zeta _{4})\alpha _{13}^{2}-
\nonumber \\
&&-(\zeta _{1}.k_{2})(\zeta _{2}.k_{3})(\zeta _{3}.k_{1})(\zeta
_{4}.k_{1})+(\zeta _{1}.k_{2})(\zeta _{2}.k_{1})(\zeta _{3}.k_{2})(\zeta
_{4}.k_{1})+(\zeta _{1}.k_{3})(\zeta _{2}.k_{1})(\zeta _{3}.k_{2})(\zeta
_{4}.k_{1})-  \nonumber \\
&&-(\zeta _{1}.k_{2})(\zeta _{2}.k_{3})(\zeta _{3}.k_{2})(\zeta
_{4}.k_{1})+(\zeta _{1}.k_{2})(\zeta _{2}.k_{1})(\zeta _{3}.k_{2})(\zeta
_{4}.k_{2})+(\zeta _{1}.k_{3})(\zeta _{2}.k_{1})(\zeta _{3}.k_{2})(\zeta
_{4}.k_{2})+  \nonumber \\
&&+(\zeta _{1}.\zeta _{4})(\zeta _{2}.k_{1})(\zeta _{3}.k_{1})\alpha _{13}+(\zeta _{1}.\zeta _{4})(\zeta _{2}.k_{1})(\zeta _{3}.k_{2})\alpha _{13}-(\zeta _{1}.k_{2})(\zeta _{2}.k_{1})(\zeta _{3}.\zeta _{4})\alpha _{13}-  \nonumber \\
&&-(\zeta _{1}.k_{3})(\zeta _{2}.k_{1})(\zeta _{3}.\zeta _{4})\alpha _{13}+(\zeta _{1}.k_{2})(\zeta _{2}.k_{3})(\zeta _{3}.\zeta _{4})\alpha _{13}+(\zeta _{1}.k_{2})(\zeta _{2}.\zeta _{3})(\zeta _{4}.k_{1})\alpha _{13}-  \nonumber \\
&&-(\zeta _{1}.\zeta _{2})(\zeta _{3}.k_{1})(\zeta _{4}.k_{1})\alpha _{13}-2(\zeta _{1}.\zeta _{2})(\zeta _{3}.k_{2})(\zeta _{4}.k_{1})\alpha _{13}+(\zeta _{1}.k_{2})(\zeta _{2}.\zeta _{3})(\zeta _{4}.k_{2})\alpha _{13} \nonumber \\
&& -(\zeta _{1}.\zeta _{2})(\zeta _{3}.k_{2})(\zeta _{4}.k_{2})\alpha _{13}
\end{eqnarray}%
\vspace{-1.4 cm}
\begin{eqnarray}
K_{5}^{(4)}&=&(\zeta _{1}.\zeta _{4})(\zeta _{2}.\zeta _{3})\alpha _{12}^{2}+(\zeta _{1}.\zeta _{2})(\zeta _{3}.\zeta _{4})\alpha _{12}^{2}+2(\zeta _{1}.\zeta _{4})(\zeta _{2}.\zeta _{3})\alpha _{13}\alpha _{12}+(\zeta _{1}.\zeta _{4})(\zeta _{2}.\zeta _{3})\alpha _{13}^{2}+  \nonumber \\
&&+(\zeta _{1}.\zeta _{4})(\zeta _{2}.k_{3})(\zeta _{3}.k_{2})\alpha _{12}-(\zeta _{3}.\zeta _{4})(\zeta _{1}.k_{2})(\zeta _{2}.k_{1})\alpha _{12}-(\zeta _{2}.\zeta _{3})(\zeta _{1}.k_{2})(\zeta _{4}.k_{1})\alpha _{12}-  \nonumber \\
&&-(\zeta _{2}.\zeta _{3})(\zeta _{1}.k_{3})(\zeta _{4}.k_{1})\alpha _{12}-(\zeta _{1}.\zeta _{2})(\zeta _{3}.k_{1})(\zeta _{4}.k_{1})\alpha _{12}-(\zeta _{1}.\zeta _{2})\alpha _{12}(\zeta _{3}.k_{2})(\zeta
_{4}.k_{1})-  \nonumber \\
&&-(\zeta _{1}.\zeta _{2})(\zeta _{3}.k_{1})(\zeta _{4}.k_{2})\alpha _{12}-(\zeta _{1}.\zeta _{2})(\zeta _{3}.k_{2})(\zeta _{4}.k_{2})\alpha _{12}+(\zeta _{1}.\zeta _{4})\alpha _{13}(\zeta _{2}.k_{3})(\zeta
_{3}.k_{2})-  \nonumber \\
&&-(\zeta _{2}.\zeta _{3})(\zeta _{1}.k_{2})(\zeta _{4}.k_{1})\alpha _{13}-(\zeta _{2}.\zeta _{3})(\zeta _{1}.k_{3})(\zeta _{4}.k_{1})\alpha _{13}+(\zeta _{1}.k_{2})(\zeta _{2}.k_{1})(\zeta _{3}.k_{1})(\zeta
_{4}.k_{1})+  \nonumber \\
&&+(\zeta _{1}.k_{2})(\zeta _{2}.k_{1})(\zeta _{3}.k_{2})(\zeta
_{4}.k_{1})-(\zeta _{1}.k_{2})(\zeta _{2}.k_{3})(\zeta _{3}.k_{2})(\zeta
_{4}.k_{1})-(\zeta _{1}.k_{3})(\zeta _{2}.k_{3})(\zeta _{3}.k_{2})(\zeta
_{4}.k_{1})+  \nonumber \\
&&+(\zeta _{1}.k_{2})(\zeta _{2}.k_{1})(\zeta _{3}.k_{1})(\zeta
_{4}.k_{2})+(\zeta _{1}.k_{2})(\zeta _{2}.k_{1})(\zeta _{3}.k_{2})(\zeta
_{4}.k_{2})
\end{eqnarray}
\vspace{-1.4 cm}
\begin{eqnarray}
K_{6}^{(4)}= &&(\zeta _{1}.\zeta _{3})(\zeta _{2}.\zeta _{4})\alpha _{13}^{2}-(\zeta _{2}.\zeta _{4})\alpha _{13}(\zeta _{1}.k_{3})(\zeta
_{3}.k_{1})+(\zeta _{1}.\zeta _{3})\alpha _{13}(\zeta _{2}.k_{1})(\zeta
_{4}.k_{2})+  \\
&&+(\zeta _{1}.\zeta _{3})\alpha _{13}(\zeta _{2}.k_{3})(\zeta
_{4}.k_{2})-(\zeta _{1}.k_{3})(\zeta _{2}.k_{1})(\zeta _{3}.k_{1})(\zeta
_{4}.k_{2})-(\zeta _{1}.k_{3})(\zeta _{2}.k_{3})(\zeta _{3}.k_{1})(\zeta
_{4}.k_{2})\nonumber
\end{eqnarray}
\vspace{-0.5cm}
\noindent Now, demanding the absence of $(\zeta .k)^{4}$ terms in the following linear
combination%

\begin{equation}
a_{3}K_{3}^{(4)}+a_{4}K_{4}^{(4)}+a_{5}K_{5}^{(4)}+a_{6}K_{6}^{(4)} \ ,
\end{equation}
\newpage
\noindent we obtain a set of 11 relations among the four unknowns $a_{3},$ $a_{4},$ $%
a_{5}$ and $a_{6}$, and we are
left with the following 3 independent equations:%
\vspace{-0.1cm}
\begin{equation}
\left\{
\begin{array}{c}
a _3+4 \ a_5=0 \\
2 \ a_4-a_3=0 \\
a_3+8 \ a_6=0%
\end{array}%
\right.
\end{equation}
\vspace{-0.2cm}
whose solution has already been shown at equation (\ref{coefficients1})
and is rewritten here
\vspace{-0.2cm}
\begin{equation}
a_3=-8 \ a_6,~~a_4=-4 \ a_6,~~a_5=2 \ a_6 \ .
\end{equation}
\vspace{-0.5cm}

\section{${\alpha'}^3$ kinematic calculations}

\label{alpha3}

In this appendix we give some details of the calculations related to the $4$ and $5$-point subamplitude of ${\cal L_{\mbox{eff}}}^{(3)}$, shown in (\ref{L3}). We do not consider the complete subamplitudes, but we will take into account only the $(\zeta . k)^n$ terms which are initially present in the corresponding $n$-point subamplitude.\\
Here we will also adopt the convention
\vspace{-0.2cm}
\begin{eqnarray}
\alpha_{ij} & = & k_i \cdot k_j \ .
\label{alphaij2}
\end{eqnarray}
\vspace{-0.2cm}
\subsection{$(\zeta \cdot k)^4$ terms}
The $(\zeta \cdot k)^4$ terms presents in the $4$-point subamplitude of ${\cal L_{\mbox{eff}}}^{(3)}$ are given by:
\vspace{-0.2cm}
\begin{eqnarray}
A_{(\zeta \cdot k)^{4}} &=&\alpha _{13}\left( (\zeta _{2}\cdot k_{1})\left\{
(\zeta _{4}\cdot k_{2})\left[ 2(\zeta _{4}\cdot k_{1})\left( \text{$a_{21}$}%
(\zeta _{1}\cdot k_{2})+(\text{$a_{16}$}-\text{$a_{17}$})(\zeta _{1}\cdot
k_{3})\right) -\right. \right. \right.  \nonumber \\
&&-\left. (\zeta _{4}\cdot k_{2})\left( 2(\text{$a_{20}$}-\text{$a_{21}$}+%
\text{$a_{22}$})(\zeta _{1}\cdot k_{2})+(2\text{$a_{17}$}+\text{$a_{22}$}%
)(\zeta _{1}\cdot k_{3})\right) \right]  \nonumber \\
&&+(\zeta _{3}\cdot k_{1})\left[ (\zeta _{4}\cdot k_{2})\left( (2(\text{$%
a_{17}$}-\text{$a_{20}$}+\text{$a_{21}$})-\text{$a_{22}$})(\zeta _{1}\cdot
k_{2})+(\text{$a_{22}$}-2(\text{$a_{17}$}+8\text{$a_{19}$}))(\zeta _{1}\cdot
k_{3})\right) \right. + \nonumber \\
&&\left. \left. +(\zeta _{4}\cdot k_{1})\left( (2(\text{$a_{17}$}+\text{$%
a_{21}$})+\text{$a_{22}$})(\zeta _{1}\cdot k_{2})+2(\text{$a_{20}$}+\text{$%
a_{22}$})(\zeta _{1}\cdot k_{3})\right) \right] \right\} + \nonumber \\
&&+(\zeta _{2}\cdot k_{3})\left\{ (\zeta _{1}\cdot k_{2})\left[ (\zeta
_{4}\cdot k_{1})\left( (2\text{$a_{17}$}+\text{$a_{22}$})(\zeta _{3}\cdot
k_{1})+2(4\text{$a_{18}$}-\text{$a_{21}$})(\zeta _{4}\cdot k_{2})\right)
\right. \right. - \nonumber \\
&&-\left. (\zeta _{4}\cdot k_{2})\left( (-2\text{$a_{16}$}+2\text{$a_{20}$}+%
\text{$a_{22}$})(\zeta _{3}\cdot k_{1})+2(\text{$a_{20}$}+\text{$a_{22}$}%
)(\zeta _{4}\cdot k_{2})\right) \right] + \nonumber \\
&&+2(\zeta _{1}\cdot k_{3})\left[ (-\text{$a_{16}$}+\text{$a_{17}$}-4\text{$%
a_{18}$}+\text{$a_{21}$})(\zeta _{4}\cdot k_{1})(\zeta _{4}\cdot
k_{2})+\right. +  \nonumber \\
&&\left. \left. \left. +(-\text{$a_{16}$}+\text{$a_{17}$}+8\text{$a_{19}$}+%
\text{$a_{20}$})(\zeta _{3}\cdot k_{1})(\zeta _{4}\cdot k_{2})\right]
\right\} \right) + \nonumber \\
&&+\alpha _{12}\left( (\zeta _{2}\cdot k_{1})\left\{ (\zeta _{1}\cdot k_{2})
\left[ 8\text{$a_{18}$}(\zeta _{4}\cdot k_{2})\left( (\zeta _{4}\cdot
k_{1})+(\zeta _{4}\cdot k_{2})\right) +\right. \right. \right.  \nonumber \\
&&\left. +(\zeta _{3}\cdot k_{1})\left( (2\text{$a_{16}$}+8\text{$a_{18}$}+%
\text{$a_{22}$})(\zeta _{4}\cdot k_{1})+(2\text{$a_{16}$}+8\text{$a_{18}$}-2%
\text{$a_{20}$}-\text{$a_{22}$})(\zeta _{4}\cdot k_{2})\right) \right] + \nonumber \\
&&+(\zeta _{1}\cdot k_{3})\left[ (\zeta _{4}\cdot k_{1})\left( 2(\text{$a_{20}
$}+\text{$a_{22}$})(\zeta _{3}\cdot k_{1})+(2\text{$a_{16}$}+\text{$a_{22}$}%
)(\zeta _{4}\cdot k_{2})\right) +\right. \nonumber  \\
&&\left. \left. +(-2\text{$a_{16}$}+2\text{$a_{20}$}+\text{$a_{22}$})\left(
(\zeta _{3}\cdot k_{1})+(\zeta _{4}\cdot k_{2})\right) (\zeta _{4}\cdot
k_{2})\right] \right\} + \nonumber \\
&&+(\zeta _{2}\cdot k_{3})\left\{ (\zeta _{4}\cdot k_{2})\left[ 2(\text{$%
a_{20}$}+\text{$a_{22}$})(\zeta _{4}\cdot k_{2})(\zeta _{1}\cdot
k_{3})+\right. \right.  \nonumber \\
&&\left. +(\zeta _{4}\cdot k_{1})\left( 8\text{$a_{18}$}(\zeta _{1}\cdot
k_{2})+(2\text{$a_{16}$}+8\text{$a_{18}$}+\text{$a_{22}$})(\zeta _{1}\cdot
k_{3})\right) \right] + \nonumber \\
&&+(\zeta _{3}\cdot k_{1})\left[ (\zeta _{4}\cdot k_{2})\left( (4\text{$a_{16}
$}-2\text{$a_{20}$})(\zeta _{1}\cdot k_{2})+(2\text{$a_{16}$}+\text{$a_{22}$}%
)(\zeta _{1}\cdot k_{3})\right) \right. + \nonumber \\
&&\left. \left. \left. +(\zeta _{4}\cdot k_{1})\left( (2\text{$a_{16}$}+%
\text{$a_{22}$})(\zeta _{1}\cdot k_{2})+2(\text{$a_{20}$}+\text{$a_{22}$}%
)(\zeta _{1}\cdot k_{3})\right) \right] \right\} \right)
\end{eqnarray}
\vspace{-0.2cm}
Now, we demand that $A_{(\zeta \cdot k)^{4}} = 0$, which implies a set of 29 relations among the seven coefficients $a_{16}$, $a_{17}$, $a_{18}$, $a_{19}$, $a_{20}$, $a_{21}$ and $a_{22}$. However, we have only 6 independent relations, resulting in
\vspace{-0.2cm}
\begin{equation}
-2a_{16}=-2a_{17}=8a_{19}=-a_{20}=a_{22} \ \ \text{ and } \ \ \ a_{18}=a_{21}=0 \ ,
\label{Zk4}
\end{equation}
which has already been shown at expression (\ref{coefficients2}).

\subsection{$(\zeta \cdot k)^5$ terms}

Let us consider the $(\zeta \cdot k)^5$ terms present in the 5-point subamplitude of ${\cal L_{\mbox{eff}}}^{(3)}$.
In this case, we have to deal with an amplitude containing poles, which appears due to the presence of 4-leg vertices produced by the terms $D^{2}F^{4}$ in ${\cal L_{\mbox{eff}}}^{(3)}$. Therefore, we can separate the contributions $(\zeta \cdot k)^5$ into two kinds of terms:
\begin{equation}
A_{(\zeta \cdot  k)^{5}}={\cal A}_{(\zeta \cdot k)^{5}}+{\cal A}^{(poles)}_{(\zeta \cdot k)^{5}}
\end{equation}
where ${\cal A}^{(poles)}_{(\zeta \cdot k)^{5}}$ and ${\cal A}_{(\zeta \cdot k)^{5}}$ denote the terms with and without poles in the $(\zeta \cdot k)^5$ contribution to the 5-point subamplitude, respectively.
In a first step, we demand that
\begin{equation}
{\cal A}^{(poles)}_{(\zeta \cdot k)^{5}}=0
\end{equation}
and we obtain a set of 706 equations for the coefficients $a_{16},...,a_{22}$. Solving this system, we find the following relations among the coefficients:
\begin{equation}
-2a_{16}=-2a_{17}=8a_{19}=-a_{20}=a_{22} \ \ \ \text{ and } \ \ a_{18}=a_{21}=0 \ .
\label{zetak4}
\end{equation}
This is exactly the same result that we obtained when we demanded the absence of $(\zeta \cdot k)^4$ terms in the 4-point subamplitude, eq (\ref{Zk4}).
Thus, requiring the absence of poles  in the $(\zeta \cdot k)^5$ terms which are present in the 5-point subamplitude is equivalent to demanding the absence of $(\zeta \cdot k)^4$ terms in the 4-point subamplitude. Therefore, at ${\alpha'}^{3}$ order, we just need to work with the 5-point subamplitude. So, including the information about the absence of poles, eq (\ref{zetak4}), and requiring the elimination of $(\zeta \cdot k)^5$ terms, still present in the 5-point subamplitude, we obtain a set of 95 relations among the remaining 6 coefficients $a_{10},...,a_{15}$ and the coefficient  $a_{22}$. Considering only the 6 linearly independent equations, we obtain:
\begin{equation}
a_{11}=a_{13}=-2a_{15}=-ia_{22} \ \ \ \text{ and } \ \ a_{10}=a_{12}=a_{14}=0 \ ,
\label{coezetak4}
\end{equation}
which is the result already presented in the main body of this work, in eq.(\ref{coefficients3}).

\section{Some details of the ${\alpha'}^4$ calculations of ${\cal L}_{\mbox{eff}}$}

\label{Some details}

Here we present some details about the derivation of the Lagrangian shown in
eq. (\ref{L4}).

\subsection{Determining the $F^6$, $D^2F^5$ and the $D^4F^4$ terms and their coefficients}

\label{Determining}

\begin{figure}[t]
\centerline{\includegraphics*[scale=0.7,angle=0]{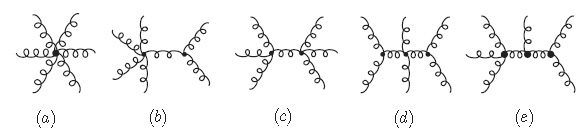}}
\caption{Feynman diagrams associated to the 6-point amplitude at ${\alpha'}^4$ order.}
\label{diagrams}
\end{figure}

It is known that using the fact that the gauge field matrices are in the adjoint representation (\ref{adjointm}), integration by parts, the $[ D , D ] \cdot = -i [F, \cdot]$ relation (\ref{group4}), Bianchi identities (\ref{Bianchi}), then a basis of terms can be found for the ${\alpha'}^4$ order terms, which has the general form
\begin{eqnarray}
{\cal L}^{(4)}_{\mbox{eff}}=(2\alpha ^{\prime })^{4}\mbox{tr}\Big[\
\sum_{i=1}^{m_{1}}a_{i}\ (F^{6})_{i}+\sum_{i=1}^{m_{2}}b_{i}\
(D^{2}F^{5})_{i}+\sum_{i=1}^{m_{3}}c_{i}\ (D^{4}F^{4})_{i}\ \Big] \ .
\label{generalL4}
\end{eqnarray}
From \cite{Koerber4} we know that this basis of terms should be 96-dimensional, that is, it should happen that $m_1+m_2+m_3=96$ but, as we will see in the next lines, based on previously known expressions for the $D^2F^5$ and the $D^4F^4$ terms, we will work with and ${\alpha'}^4$ expression which contains only 53 unknowns.\\
The explicit form of the $\{ F^6, D^2 F^5, D^4 F^4 \}$ terms that we have used in our calculations is the following:
\begin{enumerate}
\item{ \emph{$F^6$ terms:}}
We have constructed them using the tensor structure
$t_{(12)}^{\mu_1 \nu_1 \mu_2 \nu_2 \mu_3 \nu_3 \mu_4 \nu_4 \mu_5 \nu_5 \mu_6 \nu_6} \\ \mbox{tr}(F_{\mu_1 \nu_1}F_{\mu_2 \nu_2}F_{\mu_3 \nu_3} F_{\mu_4 \nu_4} F_{\mu_5 \nu_5}  F_{\mu_6 \nu_6})$, where the $t_{(12)}$ tensor is antisymmetric on each pair $(\mu_j \ \nu_j)$ and also it satisfies the relation
\begin{eqnarray}
t_{(12)}^{\mu_1 \nu_1 \mu_2 \nu_2 \mu_3 \nu_3 \mu_4 \nu_4 \mu_5 \nu_5 \mu_6 \nu_6} = + \ t_{(12)}^{\mu_1 \nu_1 \mu_6 \nu_6 \mu_5 \nu_5 \mu_4 \nu_4 \mu_3 \nu_3 \mu_2 \nu_2} \ ,
\label{t12F6}
\end{eqnarray}
which corresponds to demanding invariance of the $6$-point amplitude under a world-sheet parity transformation (`twisting' on the disk with respect to index `1').\\
The resulting $t_{(12)}$ tensor contains 28 free coefficients, that is, there are 28 independent $F^6$ terms ($m_1 = 28$ in (\ref{generalL4})).
\item{\emph{$D^2F^5$ and $D^4F^4$ terms:}}For these terms we first recall that we have explicitly determined them (and their coefficients) in eqs. (5.16) and (5.17) of ref. \cite{Barreiro1}, although in that reference we did not worry in writing them in a reduced way (and we have now confirmed that, in fact, this happens). So, in (\ref{generalL4}) we will use as an Ansatz the complete list of groupped $D^2F^5$ and $D^4F^4$ terms of ref. \cite{Barreiro1}, while keeping their $b_i$ and $c_i$ coefficients free. By `groupped' we mean that for the $D^2F^5$ terms we have used a list of 15 terms\footnote{See formula (5.16)  of ref. \cite{Barreiro1} for more details.} like $t_{(10)}^{\mu_1 \nu_1 \mu_2 \nu_2 \mu_3 \nu_3 \mu_4 \nu_4 \mu_5 \nu_5} \mbox{tr}(F_{\mu_1 \nu_1}F_{\mu_2 \nu_2}F_{\mu_3 \nu_3}D^{\alpha}F_{\mu_4 \nu_4}D_{\alpha}F_{\mu_5 \nu_5} )$, ${(\eta \cdot t_{(8)})}^{\mu_1 \nu_1 \mu_2 \nu_2 \mu_3 \nu_3 \mu_4 \nu_4 \mu_5 \nu_5} \mbox{tr} ( F_{\mu_1 \nu_1}F_{\mu_2 \nu_2}D^{\alpha}F_{\mu_3 \nu_3}D_{\alpha}F_{\mu_4 \nu_4}F_{\mu_5 \nu_5})$ and $t_{(8)}^{\mu_4 \nu_4 \mu_5 \nu_5 \mu_1 \nu_1 \mu_2 \nu_2} \\ \mbox{tr} ( D^{\mu_3}F_{\mu_1 \nu_1}D^{\nu_3}F_{\mu_2 \nu_2}F_{\mu_3 \nu_3}F_{\mu_4 \nu_4}F_{\mu_5 \nu_5})$. \\
    Now, in the case of the $D^4F^4$ terms of ref. \cite{Barreiro1} there is a subtlety: there we wrote them as a group of 8 terms like $t_{(8)}^{\mu_1 \nu_1 \mu_2 \nu_2 \mu_3 \nu_3 \mu_4 \nu_4} \mbox{tr} ( D^2 D^2F_{\mu_1 \nu_1}F_{\mu_2 \nu_2}F_{\mu_3 \nu_3}F_{\mu_4 \nu_4} )$, $t_{(8)}^{\mu_1 \nu_1 \mu_2 \nu_2 \mu_3 \nu_3 \mu_4 \nu_4} \mbox{tr} ( D^{\alpha}F_{\mu_1 \nu_1}D_{\alpha}F_{\mu_2 \nu_2}D^{\beta}F_{\mu_3 \nu_3}D_{\beta}F_{\mu_4 \nu_4} )$ and $t_{(8)}^{\mu_1 \nu_1 \mu_2 \nu_2 \mu_3 \nu_3 \mu_4 \nu_4} \\ \mbox{tr}( D^{2}F_{\mu_1 \nu_1}F_{\mu_2 \nu_2}D^{\beta}F_{\mu_3 \nu_3}D_{\beta}F_{\mu_4 \nu_4} )$. In that list of 8 terms only 2 of them do not contain a `quadratic' covariant derivative $D^2=D^{\alpha} D_{\alpha}$. The 6 remaining ones can be rewritten in terms of $F^6$ and $D^2F^5$ terms after using the identity
    \begin{eqnarray}
    D^2F_{\mu \nu} =  D^{\alpha}(D_{\alpha} F_{\mu \nu}) =
    D_{\mu}(D_{\alpha} {F^{\alpha}}_{\nu}) - D_{\nu}(D_{\alpha}
    {F^{\alpha}}_{\mu}) + 2 \ i \ [ F_{\alpha \mu},
    {F_{\nu}}^{\alpha} ] \ . \label{D2}
    \end{eqnarray}
    In fact, in the next subsection \ref{Identities} we have written explicitly various of these relations.
    At the end, we are left just with two type of $D^4F^4$ terms which do not contain `quadratic' covariant derivatives:
    \begin{eqnarray}
     t_{(8)}^{\mu_1 \nu_1 \mu_2 \nu_2 \mu_3 \nu_3 \mu_4 \nu_4} \ \tr ( D^{\alpha}F_{\mu_1 \nu_1}D_{\alpha}F_{\mu_2 \nu_2}D^{\beta}F_{\mu_3 \nu_3}D_{\beta}F_{\mu_4 \nu_4} )     \ \ \  \mbox{and} \ \nonumber \\
      t_{(8)}^{\mu_1 \nu_1 \mu_2 \nu_2 \mu_3 \nu_3 \mu_4 \nu_4} \ \tr ( D^{\alpha}F_{\mu_1 \nu_1}D_{(\beta} D_{\alpha)}F_{\mu_2 \nu_2}D^{\beta}F_{\mu_3 \nu_3}F_{\mu_4 \nu_4} ) \ .
      \label{twoterms}
      \end{eqnarray}
\end{enumerate}
\noindent Now, with respect to the determination of the coefficientes $\{ a_i, b_i, c_i\}$ in eq. (\ref{generalL4}), which we have just seen that consists in a list of $28+15+8= 51$ numbers at all, we will determine them consistently using our revisited S-matrix method.\\
\noindent First of all, demanding absence of $(\zeta \cdot k)^4$ terms in the 4-point subamplitude leads to no restrictions for the $D^4F^4$ terms because the two that appear in (\ref{twoterms}) satisfy this requirement independently.\\
 Then, when we demand absence of terms $(\zeta \cdot k)^5$ in the
5-point subamplitude of ${\cal L}^{(4)}_{\mbox{eff}}$, this procedure leads to a set of 471 non zero relations among the $b_{i}$'s and the $c_{i}$'s, from which only 8 are linearly independent.\\
\noindent After that, we require the absence of $(\zeta \cdot k)^6$ terms in the 6-point subamplitude. This procedure leads to a set of 2511 non zero relations among the $a_{i}$'s, the $b_{i}$'s, the $c_{i}$'s and $d_{1}$, from which 33 are linearly independent. Combining these 33 relations with those 8 independent relations coming from the 5-point subamplitude, we have reached a set of 41 independent relations. At this step there is an important subtlety that has to do with the fact that the gauge group matrices $\lambda^a$'s are in the $adjoint$ representation (see eq.(\ref{adjointm})). This implies that in the expression for the complete amplitude ${\cal A}_6$ (see in eq. (\ref{N-point})) there are not $(6-1)!=120$ terms, but only $(6-1)!/2=60$ ones, where the coefficient of $\mbox{tr}(\lambda^{a1} \lambda^{a2}\lambda^{a3}\lambda^{a4}\lambda^{a5}\lambda^{a6} )$ is not just $A(1,2,3,4,5,6)$, but $A(1,2,3,4,5,6)+A(6,5,4,3,2,1)$. So the absence of $(\zeta \cdot k)^6$ terms should be demanded in $A(1,2,3,4,5,6)+A(6,5,4,3,2,1)$ and not only in $A(1,2,3,4,5,6)$\footnote{Stricitly speaking, this care should have also be taken with the absence of $(\zeta \cdot k)^4$ and $(\zeta \cdot k)^5$ terms, in the $4$ and $5$-point subamplitudes, respectively, but we have checked that in those cases it has not made any difference at all in the system of equations for the coefficients.}.\\
\noindent Our result is that, after this procedure is done, we arrive to a non homogeneous system of linear equations for the $\{ a_i, b_i, c_i\}$'s which is consistently overdetermined. With it and the one that we obtained before from the absence of $(\zeta \cdot k)^5$ terms, we are able to find 39 coefficients and the remaining 12 ones remain completely arbitrary. After substituing this solution in (\ref{generalL4}) the contributions of the 12 still arbitrary coefficients dissappear $completely$ due to identities like
\begin{eqnarray}
\mbox{tr} \big( F_{\mu_1 \nu_1} D^{\mu_1}  F_{\mu_2 \nu_2}   F_{\mu_3 \nu_3}   F_{\mu_4 \nu_4} D^{\nu_1}  F_{\mu_5 \nu_5}   \big)  t_{(8)}^{\mu_2 \nu_2 \mu_3 \nu_3 \mu_4 \nu_4 \mu_5 \nu_5}= \hspace{4cm}  \nonumber \\
- \ \mbox{tr} \big(  D^{\mu_2} F_{\mu_1 \nu_1}  F_{\mu_2 \nu_2}  D^{\nu_2} F_{\mu_3 \nu_3}   F_{\mu_4 \nu_4}  F_{\mu_5 \nu_5}   \big)  t_{(8)}^{\mu_3 \nu_3 \mu_4 \nu_4 \mu_5 \nu_5 \mu_1 \nu_1}
\label{i1}
\end{eqnarray}
and
\begin{eqnarray}
\mbox{tr} \big( F_{\mu_1 \nu_1}  F_{\mu_2 \nu_2}   F_{\mu_3 \nu_3}\hspace{-4pt} &D^{\alpha}&\hspace{-4pt}  F_{\mu_4 \nu_4} D_{\alpha}  F_{\mu_5 \nu_5}   \big)  t_{(10)}^{\mu_1 \nu_1 \mu_2 \nu_2 \mu_3 \nu_3 \mu_4 \nu_4 \mu_5 \nu_5}= \hspace{4cm}  \nonumber \\
&+& \hspace{-4pt} \mbox{tr} \big(  F_{\mu_1 \nu_1}  F_{\mu_2 \nu_2} D^{\alpha}  F_{\mu_3 \nu_3}  D_{\alpha}  F_{\mu_4 \nu_4}  F_{\mu_5 \nu_5}   \big)  t_{(10)}^{\mu_1 \nu_1 \mu_2 \nu_2 \mu_3 \nu_3 \mu_4 \nu_4 \mu_5 \nu_5}
\label{i2}
\end{eqnarray}
and more complicated ones (which, for simplicity, we have not written down here).\\
\noindent So, the conclusion is that, working with our Ansatz for the $D^2 F^5$ and $D^4F^4$ terms, our revisited S-matrix method leads to $unique$ and completely determined lagrangian at ${\alpha'}^4$, ${\cal L}^{(4)}_{\mbox{eff}}$, given in eq.(\ref{L4}) (together with the auxiliary equations (\ref{LF6}),  (\ref{LD2F5}) and (\ref{LD4F4})) of the main body of this work. At the end, there is no arbitrary coefficient in ${\cal L}^{(4)}_{\mbox{eff}}$. \\
In the rest of this section of Appendix \ref{Some details} we give some further details about demanding the absence of $(\zeta \cdot k)^6$ in $A(1,2,3,4,5,6)+A(6,5,4,3,2,1)$ at order ${\alpha'}^4$.\\
\noindent First of all, in Figure \ref{diagrams} we have drawn all type of 6-point tree level Feynman diagrams that arise at ${\alpha'}^4$ order.\\
\noindent {\bf i)} We have checked that the diagrams with two propagators ($(d)$ and $(e)$ in Figure \ref{diagrams}) do not contain $(\zeta \cdot k)^6$ terms.\\
\noindent {\bf ii)} We have checked that the relations among the $b_{i}$'s and the $c_{i}$'s obtained by demanding absence of terms $(\zeta \cdot k)^5$ automatically avoid the presence of simple poles on the $(\zeta \cdot k)^6$ terms of $A(1,2,3,4,5,6)+A(6,5,4,3,2,1)$, that is, the simple pole contributions to the $(\zeta \cdot k)^6$ terms of diagrams $(b)$, $(c)$, $(d)$ and $(e)$ of Figure \ref{diagrams} is zero (after using momentum conservation, the physical state and the on-shell conditions). \\
\noindent {\bf iii)} Let us analyze more carefully the diagram $(c)$ in Figure \ref{diagrams}. That diagram has contributions from vertices coming $\mathcal{L}_{YM}+(2\alpha ^{\prime })^{4}%
\mathcal{L}_{D^{4}F^{4}}^{(4)}$, but it also has contributions from 4-point vertices that come from  $(2\alpha ^{\prime })^{2}\mathcal{L}%
_{F^{4}}^{(2)}$, presented in eq. (\ref{L12simpler12})\footnote{It is not obvious to see, but it happens that the $(\zeta \cdot k)^6$ terms coming from the contributions of diagrams $(c)$ constructed with these vertices are terms which have no poles at all (after using momentum conservation, the physical state and the on-shell conditions).}. This is a crucial point in our
calculations. It is due to the contribution of these last terms that the linear system for the $\{ a_i, b_i, c_i\}$'s is non homogeneous. This system arises from demanding absence of $(\zeta \cdot k)^6$ in the contribution without poles in $A(1,2,3,4,5,6)+A(6,5,4,3,2,1)$.\\
\noindent {\bf iv)} We have checked that, within the scheme that we have proposed as basis for the $D^2 F^5$ and the $D^4F^4$ terms, in order to arrive to the final expression for ${\cal L}^{(4)}_{\mbox{eff}}$ we have needed to use the information of both, absence of $(\zeta \cdot k)^5$ terms and absence of $(\zeta \cdot k)^6$ terms. If we only used the information that comes from demanding absence of $(\zeta \cdot k)^6$ terms, this is not enough to arrive to our final solution: there would still be some undetermined coefficients in ${\cal L}^{(4)}_{\mbox{eff}}$. \\
We are not sure what might have happened if we had found a 96-dimensional basis of $F^6$, $D^2F^5$ and $D^4F^4$ terms and then only demanded the absence of $(\zeta \cdot k)^6$ terms: may be that could have been a sufficient requirement to determine ${\cal L}^{(4)}_{\mbox{eff}}$ completely, in the same way as demanding absence of $(\zeta \cdot k)^5$ terms in the $5$-point amplitude was enough to determine ${\cal L}^{(3)}_{\mbox{eff}}$ completely (see the final part of subsection \ref{Low3} in the main body of this work).

\subsection{Identities involving quadratic covariant derivatives}

\label{Identities}

In this section of Appendix \ref{Some details} we only mention some of the identities that we refered to in section \ref{Determining}. \\
Using the identity in (\ref{D2}) and disconsidering the terms containing $D_{\alpha}F^{\alpha \mu}$ (which do not contribute in the on-shell scattering amplitudes) it can be proved that
\begin{eqnarray}
\label{ident1}
t_{(8)}^{\mu _{1} \nu _{1} \mu _{2} \nu _{2} \mu _{3} \nu _{3} \mu _{4} \nu
_{4}}&\text{tr}&\left( D_{\beta }F_{\mu _{1}\nu _{1}}D^{(2} D^{\beta )}F_{\mu
_{2}\nu _{2}}F_{\mu _{3}\nu _{3}}F_{\mu _{4}\nu _{4}}\right) =  \nonumber \\
&-&i \ \text{tr}\left( F_{\mu _{1}\nu _{1}}D^{\mu _{1}}F_{\mu _{2}\nu _{2}}F_{\mu
_{3}\nu _{3}}F_{\mu _{4}\nu _{4}}D^{\nu _{1}}F_{\mu _{5}\nu _{5}}\right)
t_{(8)}^{\mu _{2} \nu _{2} \mu _{3} \nu _{3} \mu _{4} \nu _{4} \mu _{5} \nu_{5}} \nonumber \\
&&-i \ \text{tr}\left( D^{\mu _{3}}F_{\mu _{1}\nu _{1}}D^{\nu _{3}}F_{\mu
_{2}\nu _{2}}F_{\mu _{3}\nu _{3}}F_{\mu _{4}\nu _{4}}F_{\mu _{5}\nu
_{5}}\right) t_{(8)}^{\mu _{4} \nu _{4} \mu _{5} \nu _{5} \mu _{1} \nu_{1} \mu _{2} \nu _{2}} \nonumber \\
&&-i \ (\eta \cdot t_{(8)})^{\mu _{1} \nu _{1}\mu _{2}\nu _{2}\mu_{3} \nu _{3} \mu _{4} \nu _{4} \mu _{5} \nu _{5}}\text{tr}\left( F_{\mu
_{1}\nu _{1}}D_{\alpha }F_{\mu _{2}\nu _{2}}D^{\alpha }F_{\mu _{3}\nu
_{3}}F_{\mu _{4}\nu _{4}}F_{\mu _{5}\nu _{5}}\right)  \nonumber \\
&&-i \ (\eta \cdot t_{(8)})^{\mu _{1} \nu _{1} \mu _{2} \nu _{2} \mu
_{3} \nu _{3} \mu _{4} \nu _{4} \mu _{5} \nu _{5}}\text{tr}\left( F_{\mu
_{1}\nu _{1}}D_{\alpha }F_{\mu _{2}\nu _{2}}F_{\mu _{3}\nu _{3}}D^{\alpha
}F_{\mu _{4}\nu _{4}}F_{\mu _{5}\nu _{5}}\right) \ , \\
\label{ident2}
t_{(8)}^{\mu _{1} \nu _{1} \mu _{2} \nu _{2} \mu _{3} \nu _{3} \mu _{4} \nu_{4}}&\text{tr}&\left( D^{2}D^{2}F_{\mu _{1}\nu _{1}}F_{\mu _{2}\nu
_{2}}F_{\mu _{3}\nu _{3}}F_{\mu _{4}\nu _{4}}\right) = \nonumber  \\
&-&2i \ (\eta \cdot t_{(8)})^{\mu _{1} \nu _{1} \mu _{2} \nu _{2} \mu _{3} \nu _{3} \mu_{4} \nu _{4} \mu _{5} \nu _{5}}\text{tr}\left( F_{\mu _{1}\nu _{1}}F_{\mu
_{2}\nu _{2}}D_{\alpha }F_{\mu _{3}\nu _{3}}D^{\alpha }F_{\mu _{4}\nu
_{4}}F_{\mu _{5}\nu _{5}}\right)  \nonumber \\
&&-2 \ {(\eta \cdot \eta \cdot t_{(8)})_{1}}^{\mu _{1} \nu_{1} \mu _{2} \nu _{2} \mu _{3} \nu _{3} \mu _{4} \nu _{4} \mu _{5} \nu_{5} \mu _{6} \nu _{6}}\text{tr}\left( F_{\mu _{1}\nu _{1}}F_{\mu _{2}\nu
_{2}}F_{\mu _{3}\nu _{3}}F_{\mu _{4}\nu _{4}}F_{\mu _{5}\nu _{5}}F_{\mu
_{6}\nu _{6}}\right) \nonumber \\
&&-2 \ {(\eta \cdot \eta \cdot t_{(8)})_{2}}^{\mu _{1} \nu_{1} \mu _{2} \nu _{2} \mu _{3} \nu _{3} \mu _{4} \nu _{4} \mu _{5} \nu_{5} \mu _{6} \nu _{6}}\text{tr}\left( F_{\mu _{1}\nu _{1}}F_{\mu _{2}\nu
_{2}}F_{\mu _{3}\nu _{3}}F_{\mu _{4}\nu _{4}}F_{\mu _{5}\nu _{5}}F_{\mu
_{6}\nu _{6}}\right) \ , \nonumber\\
\\
\label{ident3}
t_{(8)}^{\mu _{1}\nu _{1}\mu _{2}\nu _{2}\mu _{3}\nu _{3}\mu _{4}\nu
_{4}}&\text{tr}&\left( D^{2}F_{\mu _{1}\nu _{1}}F_{\mu _{2}\nu _{2}}D^{\beta
}F_{\mu _{3}\nu _{3}}D_{\beta }F_{\mu _{4}\nu _{4}}\right) = \nonumber
\\ &-&i \ (\eta \cdot t_{(8)})^{\mu _{1}\nu _{1}\mu _{2}\nu _{2}\mu _{3}\nu _{3}\mu
_{4}\nu _{4}\mu _{5}\nu _{5}}\text{tr}\left( D^{\beta }F_{\mu _{1}\nu
_{1}}D_{\beta }F_{\mu _{2}\nu _{2}}F_{\mu _{3}\nu _{3}}F_{\mu _{4}\nu
_{4}}F_{\mu _{5}\nu _{5}}\right) \ , \\
\label{ident4}
t_{(8)}^{\mu _{1}\nu _{1}\mu _{2}\nu _{2}\mu _{3}\nu _{3}\mu _{4}\nu
_{4}}&\text{tr}&\left( D^{2}F_{\mu _{1}\nu _{1}}D^{2}F_{\mu _{2}\nu
_{2}}F_{\mu _{3}\nu _{3}}F_{\mu _{4}\nu _{4}}\right) = \nonumber \\
&& {(\eta \cdot \eta \cdot t_{(8)})_{3}}^{\mu _{1}\nu _{1}\mu _{2}\nu _{2}\mu
_{3}\nu _{3}\mu _{4}\nu _{4}\mu _{5}\nu _{5}\mu _{6}\nu _{6}}\text{tr}%
\left( F_{\mu _{1}\nu _{1}}F_{\mu _{2}\nu _{2}}F_{\mu _{3}\nu _{3}}F_{\mu
_{4}\nu _{4}}F_{\mu _{5}\nu _{5}}F_{\mu _{6}\nu _{6}}\right) \ , \\
\label{ident5}
t_{(8)}^{\mu _{1}\nu _{1}\mu _{2}\nu _{2}\mu _{3}\nu _{3}\mu _{4}\nu
_{4}}&\text{tr}&\left( D^{2}F_{\mu _{1}\nu _{1}}F_{\mu _{2}\nu
_{2}}D^{2}F_{\mu _{3}\nu _{3}}F_{\mu _{4}\nu _{4}}\right) = \nonumber \\
&& {(\eta \cdot \eta \cdot t_{(8)})_{4}}^{\mu _{1}\nu _{1}\mu _{2}\nu
_{2}\mu _{3}\nu _{3}\mu _{4}\nu _{4}\mu _{5}\nu _{5}\mu _{6}\nu _{6}}%
\text{tr}\left( F_{\mu _{1}\nu _{1}}F_{\mu _{2}\nu _{2}}F_{\mu _{3}\nu
_{3}}F_{\mu _{4}\nu _{4}}F_{\mu _{5}\nu _{5}}F_{\mu _{6}\nu _{6}}\right) \ ,
\end{eqnarray}
where the $(\eta \cdot \eta \cdot t_{(8)})_j$ 12-index tensors are defined by
\begin{eqnarray}
\label{etaetat81}
{(\eta \cdot \eta \cdot t_{(8)})_{1}}^{\mu _{1}\nu _{1}\mu _{2}\nu
_{2}\mu _{3}\nu _{3}\mu _{4}\nu _{4}\mu _{5}\nu _{5}\mu _{6}\nu
_{6}} &=&\eta ^{\mu _{1}\nu _{5}}\eta ^{\nu _{1}\nu _{6}} t_{(8)}^{\mu
_{5}\mu _{6}\mu _{2}\nu _{2}\mu _{3}\nu _{3}\mu _{4}\nu _{4}}+ \nonumber \\
&&+\left(
\begin{array}{c}
\text{7 terms coming from antisymmetrization} \\
\text{on }(\mu _{1}\nu _{1})~(\mu _{5}\nu _{5})\text{ and }(\mu _{6}\nu
_{6})%
\end{array}%
\right) \ , \\
\label{etaetat82}
{(\eta \cdot \eta \cdot t_{(8)})_{2}}^{\mu _{1}\nu _{1}\mu _{2}\nu
_{2}\mu _{3}\nu _{3}\mu _{4}\nu _{4}\mu _{5}\nu _{5}\mu _{6}\nu
_{6}} &=&{\eta}^{\mu _{1}\nu _{6}}{\eta}^{\nu _{5}\mu _{6}} t_{(8)}^{\mu
_{5}\nu _{1}\mu _{2}\nu _{2}\mu _{3}\nu _{3}\mu _{4}\nu _{4}}+ \nonumber \\
&&+\left(
\begin{array}{c}
\text{7 terms coming from antisymmetrization} \\
\text{on }(\mu _{1}\nu _{1})~(\mu _{5}\nu _{5})\text{ and }(\mu _{6}\nu
_{6})%
\end{array}%
\right) \ , \\
\label{etaetat83}
{(\eta \cdot \eta \cdot t_{(8)})_{3}}^{\mu _{1}\nu _{1}\mu _{2}\nu
_{2}\mu _{3}\nu _{3}\mu _{4}\nu _{4}\mu _{5}\nu _{5}\mu _{6}\nu
_{6}} &=&\eta^{\mu _{5}\nu _{6}}\eta^{\mu _{1}\mu _{2}} t_{(8)}^{\nu
_{5}\mu _{6}\nu _{1}\nu _{2}\mu _{3}\nu _{3}\mu _{4}\nu _{4}}+ \nonumber \\
&&+\left(
\begin{array}{c}
\text{15 terms coming from antisymmetrization} \\
\text{on }(\mu _{1}\nu _{1})~(\mu _{2}\nu _{2})~(\mu _{5}\nu _{5})\text{
and }(\mu _{6}\nu _{6})%
\end{array}%
\right) \ , \ \ \ \\
\label{etaetat84}
{(\eta \cdot \eta \cdot t_{(8)})_{4}}^{\mu _{1}\nu _{1}\mu _{2}\nu
_{2}\mu _{3}\nu _{3}\mu _{4}\nu _{4}\mu _{5}\nu _{5}\mu _{6}\nu
_{6}} &=&\eta^{\mu _{5}\mu _{6}}\eta^{\mu _{2}\mu _{3}} t_{(8)}^{\nu
_{5}\nu _{6}\mu _{1}\nu _{1}\nu _{2}\nu _{3}\mu _{4}\nu _{4}}+ \nonumber \\
&&+\left(
\begin{array}{c}
\text{15 terms coming from antisymmetrization} \\
\text{on }(\mu _{2}\nu _{2})~(\mu _{3}\nu _{3})~(\mu _{5}\nu _{5})\text{
and }(\mu _{6}\nu _{6})%
\end{array}%
\right) \ .
\end{eqnarray}
\noindent All these $(\eta \cdot \eta \cdot t_{(8)})_j$ 12-index tensors can be expressed as linear combinations of the 28 $t_{(12)}$ 12-index tensors mentioned in the previous subsection \ref{Determining}, in the construction of the $F^6$ terms (item 1).

\subsection{Tests for the ${\alpha'}^4$ terms}

\label{Tests}

\noindent {\bf 1.} 4-point amplitude:\\

\noindent We have checked that the 4-point amplitude, obtained from the two $D^4F^4$ terms mentioned in eq. (\ref{twoterms}), reproduces exactly (after considering the coefficients that have been found for ${\cal L}^{(4)}_{\mbox{eff}}$ ) the result from the open supertring 4-point amplitude, eq.(\ref{A4fermionic}) of the main body of this work, at ${\alpha'}^4$ order.\\

\noindent {\bf 2.} 5-point amplitude:\\

\noindent In our paper \cite{Barreiro1} we obtained that the open superstring 5-point amplitude could be simplified to
\begin{eqnarray}
A(1,2,3,4,5)=T\cdot A_{YM}(1,2,3,4,5)+\left( 2\alpha ^{\prime }\right)
^{2}K_{3}\cdot A_{F^{4}}(1,2,3,4,5) \ ,
\label{A12345}
\end{eqnarray}
where the ${\alpha'}^4$ contribution in the momentum factors $T$ and $(2 \alpha')^2 K_3$ is given by
\begin{eqnarray}
\left( 2\alpha ^{\prime }\right) ^{2}K_{3}=\left( 2\alpha ^{\prime }\right)
^{4}\ \frac{2}{5}\zeta (2)^{2}\ (\ I_{1}^{(2)}+\frac{1}{4}%
I_{2}^{(2)}+I_{3}^{(2)}\ ) \ , \\
T=\left( 2\alpha ^{\prime }\right) ^{4}\ \frac{2}{5}\zeta (2)^{2}\ (\
I_{8}^{(4)}+\frac{1}{4}I_{10}^{(4)}+I_{13}^{(4)}+I_{14}^{(4)}\ ) \ ,
\label{alpha4}
\end{eqnarray}
with
\begin{eqnarray}
\label{I12}
I_{1}^{(2)} &=&\alpha _{12}^{2}+\alpha _{23}^{2}+\alpha _{34}^{2}+\alpha
_{45}^{2}+\alpha _{51}^{2}\ ,   \\
I_{2}^{(2)} &=&\alpha _{12}\alpha _{23}+\alpha _{23}\alpha _{34}+\alpha
_{34}\alpha _{45}+\alpha _{45}\alpha _{51}+\alpha _{51}\alpha _{12}\ , \\
I_{3}^{(2)} &=&\alpha _{12}\alpha _{34}+\alpha _{23}\alpha _{45}+\alpha
_{34}\alpha _{51}+\alpha _{45}\alpha _{12}+\alpha _{51}\alpha _{23}\ .
\end{eqnarray}
and
\begin{eqnarray}
I_{8}^{(4)} &=&\alpha _{12}^{2}\alpha _{23}\alpha _{34}+\alpha
_{23}^{2}\alpha _{34}\alpha _{45}+\alpha _{34}^{2}\alpha _{45}\alpha
_{51}+\alpha _{45}^{2}\alpha _{51}\alpha _{12}+\alpha _{51}^{2}\alpha
_{12}\alpha _{23}\ , \\
I_{10}^{(4)} &=&\alpha _{12}^{2}\alpha _{23}\alpha _{51}+\alpha
_{23}^{2}\alpha _{34}\alpha _{12}+\alpha _{34}^{2}\alpha _{45}\alpha
_{23}+\alpha _{45}^{2}\alpha _{51}\alpha _{34}+\alpha _{51}^{2}\alpha
_{12}\alpha _{45}\ , \\
I_{13}^{(4)} &=&\alpha _{12}^{2}\alpha _{45}\alpha _{51}+\alpha
_{23}^{2}\alpha _{51}\alpha _{12}+\alpha _{34}^{2}\alpha _{12}\alpha
_{23}+\alpha _{45}^{2}\alpha _{23}\alpha _{34}+\alpha _{51}^{2}\alpha
_{34}\alpha _{45}\ , \\
\label{I144}
I_{14}^{(4)} &=&\alpha _{12}\alpha _{23}\alpha _{34}\alpha _{45}+\alpha
_{23}\alpha _{34}\alpha _{45}\alpha _{51}+\alpha _{34}\alpha _{45}\alpha
_{51}\alpha _{12}+\alpha _{45}\alpha _{51}\alpha _{12}\alpha _{23}+\alpha
_{51}\alpha _{12}\alpha _{23}\alpha _{34}\ .\ \
\end{eqnarray}
In formulas (\ref{I12}) to (\ref{I144})  we have used the convention $\alpha_{ij} = k_i \cdot k_j$.\\
We have checked that the 5-point amplitude coming from the $D^2F^5$ and the $D^4F^4$ terms agrees exactly with the one in (\ref{A12345}) at ${\alpha'}^4$ order, once we have introduce the known expressions of the $A_{YM}(1,2,3,4,5)$ and the $A_{F^{4}}(1,2,3,4,5)$ subamplitudes. \\

\noindent {\bf 3.} Abelian limit:\\

\noindent We have checked that the abelian limit of our $F^6$ terms is given by
\begin{eqnarray}
\mathcal{L}_{Abelian} &=&(2\pi \alpha ^{\prime })^{4}\Big(-\frac{1}{12}%
F_{\mu _{1}}^{\ \mu _{2}}F_{\mu _{2}}^{\ \mu _{3}} F_{\mu _{3}}^{\ \mu _{4}} F_{\mu _{4}}^{\ \mu_{5}}
 F_{\mu _{5}}^{\ \mu _{6}}F_{\mu _{6}}^{\ \mu _{1}} \ +\frac{1}{32}F_{\mu _{1}}^{\ \mu _{2}} F_{\mu _{2}}^{\ \mu _{3}} F_{\mu _{3}}^{\ \mu_{4}} F_{\mu _{4}}^{\ \mu _{1}}
F_{\mu _{5}}^{\ \mu _{6}} F_{\mu _{6}}^{\ \mu _{5}} - \nonumber \\
&&  \hphantom{ (2\pi \alpha ^{\prime })^{4}\big(  } -\frac{1}{384}F_{\mu _{1}}^{\ \mu _{2}} F_{\mu _{2}}^{\ \mu _{1}} F_{\mu _{3}}^{\ \mu_{4}} F_{\mu _{4}}^{\ \mu _{3}} F_{\mu _{5}}^{ \ \mu _{6}}F_{\mu _{6}}^{\ \mu _{5}}\Big) \ ,
\label{LF6abelian}
\end{eqnarray}
which is the correct form for this limit as seen in the literature \cite{Andreev1, Koerber2}.

%\section{Upper bound for the number of superinvariants at ${\alpha'}^p$ order}

%\label{Upper}

\end{document}